\documentclass[superscriptaddress,reprint,hidelinks,aps,pra,longbibliography]{revtex4-1}
\usepackage{graphics}
\usepackage{bm}
\usepackage{bbm}
\usepackage{amsmath}
\usepackage{amssymb}
\usepackage{graphicx}
\usepackage{amsthm}
\usepackage{hyperref}

\begin{document}

\title{Bipartite representations and many-body entanglement of  pure states of  $N$ indistinguishable particles}

\author{J.A.\ Cianciulli}
\affiliation{IFLP/CONICET--Departamento  de F\'{\i}sica,
    Universidad Nacional de La Plata, C.C. 67, La Plata (1900), Argentina}
    \author{R.\ Rossignoli}
\affiliation{IFLP/CONICET--Departamento  de F\'{\i}sica,
    Universidad Nacional de La Plata, C.C. 67, La Plata (1900), Argentina}
\affiliation{Comisi\'on de Investigaciones Cient\'{\i}ficas (CIC), La Plata (1900), Argentina}
\author{M.\ Di Tullio}
\affiliation{IFLP/CONICET--Departamento  de F\'{\i}sica,
    Universidad Nacional de La Plata, C.C. 67, La Plata (1900), Argentina}
\author{N.\ Gigena}
\affiliation{IFLP/CONICET--Departamento  de F\'{\i}sica,
    Universidad Nacional de La Plata, C.C. 67, La Plata (1900), Argentina}
    \author{F.\ Petrovich}
\affiliation{IFLP/CONICET--Departamento  de F\'{\i}sica,
    Universidad Nacional de La Plata, C.C. 67, La Plata (1900), Argentina}

\begin{abstract}
We analyze a general bipartite-like  representation of arbitrary pure states of $N$ indistinguishable particles, valid for both bosons and fermions, based on $M$- and $(N\!-\!M)$-particle states.  It leads to  exact $(M,N\!-\!M)$ Schmidt-like expansions of the state for any $M<N$ and is  directly related to the isospectral reduced  $M$- and $(N\!-\!M)$-body density matrices $\rho^{(M)}$ and $\rho^{(N-M)}$.     
The formalism also allows for reduced yet still exact Schmidt-like decompositions associated with blocks of these densities, in systems having a fixed fraction of the particles in  some single particle subspace.  Monotonicity of the ensuing $M$-body entanglement under  a certain set of quantum operations is also discussed.  
 Illustrative examples in fermionic and bosonic systems with pairing correlations are provided, which show that in the presence of dominant eigenvalues in $\rho^{(M)}$, approximations based on a few terms of the pertinent Schmidt expansion can provide a reliable description of the state.  The associated one- and two-body  entanglement spectrum and entropies are also analyzed.
\end{abstract}
\maketitle

\section{Introduction\label{I}}

Quantum entanglement and particle indistinguishability are  
undoubtedly among the most fundamental features of quantum mechanics. Yet the extension of the concept of entanglement 
 to systems of indistinguishable particles is not straightforward 
 \cite{BFFM.20}. The standard theory of entanglement  \cite{NC.00,HHHH.09} was originally devised for systems of distinguishable components, where the pertinent Hilbert space has a tensor product structure which plays 
 an essential role already in the basic definition:  separable states, i.e.\ those  that can be generated by local operations and classical communication, are just product states (or convex mixtures of product states in the mixed case), all remaining states being entangled. 

In systems of indistinguishable components  all states are, however,  necessarily symmetrized  (bosons) or antisymmetrized (fermions), preventing in principle a direct extension of previous scheme. The definition of entanglement in these systems has then followed different approaches, starting from {\it  mode entanglement} \cite{Za.02,Shi.03,AM.07,HV.0909,FL.13,SSG.18}, where each side has access to different orthogonal (and hence distinguishable) single particle (sp) modes and  entanglement is then defined in the standard form, 
   although it  becomes thus dependent on the choice of sp modes. A distinct  approach is the so-called  {\it particle entanglement} or {\it entanglement beyond symmetrization}  \cite{SC.01,ES.02,PY.01,LZL.01,GM.0204,WV.03,IV.13,GR.15,MB.16,SD.18,GDR.20,GDR.21}, 
   which is independent of the choice of basis and just basic independent particle states  (i.e., Slater determinants (SDs) for fermions) are nonentangled. Other proposals based on correlations between observables or measurements have also been discussed  \cite{BK.04,ZL.04,SI.11,BG.1313,BF.14}, including the consideration of symmetrization correlations as entanglement \cite{CM.07,KC.14,CC.18,FC.18,MY.20}. 
Connections between these distinct forms of entanglement  and their behavior in different contexts  have been analyzed by several authors   \cite{BFFM.20,IMV.15,DD.16,BF.17,GR.17,
DI.20,DS.2021,DGR.18,DRCG.19,FMR2.2122,KKS.22,GC.23,HRS.23}. 

In particular, in \cite{GR.15} we focused on the {\it one-body entanglement} in fermion systems,  which is  determined by the one-body reduced density matrix (DM) $\rho^{(1)}$ and 
 vanishes just for SDs (or quasiparticle vacua when extended to states with no fixed particle number). It also represents  the {\it minimum} mode entanglement associated with a sp basis \cite{GR.15}  and is connected 
 to the minimum bipartite mode entanglement in four-level systems \cite{GR.17}. 
  In \cite{GDR.20} we examined its interpretation as a quantum resource \cite{CG.19,PRXM.18} in fermion systems, 
  showing through a general majorization relation that it cannot decrease under a class of sp measurements, and also identified its relation with a  bipartite-like $(1,N\!-\!1)$ representation of a general $N$-fermion state. In \cite{GDR.21} we extended the previous scheme to general {\it $M$-body entanglement} in $N$-fermion states ($M<N$), determined by the $M$-body DM $\rho^{(M)}$.  
  Interest on many-body DMs and their relation with entanglement and characterization of correlations have recently increased in different areas \cite{IMV.15,Fe.22,AF.17,Samb.20,NL.23}.  
  
 The aim of this work is first to extend the formalism of \cite{GDR.21} to the bosonic case, 
 developing a unified second-quantized formalism valid for both bosons and fermions. The formalism  still conserves, nonetheless, some of the features of the standard distinguishable case: 
 Any pure state of $N$ indistinguishable particles 
 is shown to admit, for $1\leq M\leq N\!-\!1$, a bipartite-like  $(M,N\!-\!M)$ representation and Schmidt-like decomposition, whose coefficients are   independent of the choice of sp basis and are just the square roots of the eigenvalues of the reduced  DMs $\rho^{(M)}$ and $\rho^{(N-M)}$, isospectral in any pure state. The ensuing  $(M,N\!-\!M)$ entanglement  determined by the mixedness of these eigenvalues is shown, through a general majorization relation,  to be nonincreasing under a certain set of $L$-particle operations (and to stay invariant under unitary sp transformations), thus opening the way to a basic  mode-independent 
  $(M,N\!-\!M)$ quantum resource theory. 
 Besides, for $M>1$ the $M$-body DMs may exhibit a few large {\it dominant}  eigenvalues in correlated states, enabling a reliable  approximation  of the state based on just a few terms of the associated Schmidt expansion. 
 
 In addition, we also consider here the case of states  having a fixed fraction of the total number of particles within some sp subspace. This is a common situation, arising e.g.\ in  eigenstates of interacting Hamiltonians conserving the number of particles within certain sp subspaces, like a set ${\cal S}$ and a partner set $\bar{\cal S}$ of time-reversed sp states in  pairing-type Hamiltonians or Hubbard models, as discussed in sections \ref{III} and \ref{IV}. It also emerges, of course, in any bosonic or fermionic entangled state having fixed number $N_i $ of particles in orthogonal subsets ${\cal S}_i$ of modes, like e.g.\ distinct sites. 
  In such cases the $M$-body DMs exhibit a blocked structure, and it will be shown that reduced but still exact $(M,N\!-\!M)$ bipartite-like expansions and Schmidt decompositions  associated with each of these blocks are also possible. Moreover, the standard distinguishable bipartite case naturally emerges here as a particular instance in this general formulation. 

 The general formalism is presented in section \ref{II}, while the special case of blocked DMs and reduced exact expansions are treated in section \ref{III}. Illustrative examples in  finite systems with pairing correlations are provided in section \ref{IV} for both fermions and bosons. They    
 include analytical results and bounds for the one and two-body  entanglement spectrum in some typical paired states, as well as  numerical results for the latter and the associated entanglement entropy in the exact GS of a finite pairing Hamiltonian. These results show the presence of a characteristic large dominant eigenvalue in the two-body DM of both fermionic and bosonic paired states, together with a highly mixed one-body DM, which provides a clear signature of such states.  
 It is also shown that through such eigenvalue and the associated eigenvector, a good approximation to the exact GS of the previous Hamiltonian for {\it all} values of the coupling strength can be here achieved with just very few terms of the pertinent $(2,N\!-\!2)$ Schmidt expansion. Conclusions are finally provided in section \ref{V}.

\section{Formalism\label{II}}
\subsection{$N$ particle states in boson and fermion systems }
Let us consider a single particle (sp) space $\cal H$ of finite dimension $d$ and a set of particle creation 
and annihilation operators $c^\dag_i$, $c_i$,  $i=1,\ldots,d$, 
satisfying 
\begin{equation}
[c_i,c_j]_{\pm}=[c^\dag_i,c^\dag_j]_{\pm}=0\,,\;[c_i,c^\dag_j]_{\pm}=\delta_{ij}\,,
\label{cnm}
\end{equation}
for bosons ($-$) or fermions ($+$), where $[a,b]_{\pm}=ab\pm ba$.  
We define the $M$-particle  creation  operators 
\begin{equation}    C^{(M)\dag}_{\bm\alpha}=\frac{c^{\dag n_1}_1}{\sqrt{n_1!}}\frac{c^{\dag n_2}_2}{\sqrt{n_2!}}\ldots \frac{c^{\dag n_d}_d}{\sqrt{n_d!}}\,,\;\;\;\sum_{i=1}^d n_i=M\,,
    \label{Cm}    
\end{equation}
where $n_i=0,1,2,\ldots$ for bosons and 
$n_i=0,1$ for fermions, while  $\bm{\alpha}=(n_1,\ldots,n_d)$. When applied to the vacuum $|0\rangle$,  
 they create normalized orthogonal $M$-particle states $C^{(M)\dag}_{\bm\alpha}|0\rangle$ having  $n_i$ particles in sp state $i$. For $M\geq 0$ (and $M\leq d$ for fermions) these states span  the full Fock space ${\cal F}$ associated with ${\cal H}$,  satisfying    
\begin{equation}
\langle 0|C_{\bm\alpha}^{(M)}C_{\bm\alpha'}^{(M')\dag}|0\rangle=\delta^{MM'}\delta_{\bm\alpha\bm\alpha'}\,. 
\label{ort}
\end{equation}

The  subspace ${\cal F}_M$ of $M$-particle states has dimension $d_M
=\binom{d+M-1}{M}$ for bosons and $d_M
=\binom{d}{M}$ for fermions, and is generated by the $d_M$ operators \eqref{Cm}. 
These operators  satisfy   $C_{\bm\alpha}^{(M)}C^{(M)\dag}_{\bm\alpha'}|0\rangle=\delta_{\bm\alpha\bm\alpha'}|0\rangle$ and  (see App.\ \ref{A})
  \begin{eqnarray}
\sum_{\bm \alpha}C^{(M)\dag}_{\bm \alpha} 
     C^{(M)}_{\bm\alpha}&=&\binom{\hat{N}}{M}\,,\label{Nm}
\end{eqnarray}
for both fermions and bosons, where $\hat{N}=\sum_{i}c^\dag_i c_i$ is the particle number operator and $\binom{\hat N}{M}$  the operator taking the value $\binom{N}{M}=\frac{N!}{M!(N-M)!}$  when applied to an  $N$-particle state: $\binom{\hat N}{M}|\Psi\rangle=\binom{N}{M}|\Psi\rangle$ if $\hat N|\Psi\rangle=N|\Psi\rangle$. The sum over $\bm\alpha$ in \eqref{Nm} runs over all $d_M$ operators \eqref{Cm},  
 i.e.\ over all possible occupations $(n_1,\ldots,n_d)$ with $\sum_i n_i=M$.  
 For instance, 
$\sum_{\bm\alpha}C^{(2)\dag}_{\bm\alpha}C^{(2)}_{\bm\alpha}=\sum_{i<j} c^\dag_i c^\dag_j c_j c_i +\sum_i\frac{{c_i^{\dag}}^2}{\sqrt{2}}\frac{c_i^2}{\sqrt{2}}=\frac{1}{2}\sum_{i, j}c^\dag_ic^\dag_j c_j c_i
=\frac{\hat{N}(\hat{N}-1)}{2}=\binom{\hat N}{2}$ for both bosons, and fermions (where  
$i=j$ terms obviously vanish).   

An arbitrary normalized pure state $|\Psi\rangle$ of $N$ identical particles (bosons or fermions) 
 can then be written as 
\begin{subequations}
 \begin{eqnarray}
   |\Psi\rangle&=&\tfrac{1}{N!}\sum_{i_1,\ldots,i_N}\Gamma_{i_1\ldots i_N}c^\dag_{i_1}\ldots c^\dag_{i_N}|0\rangle\label{st1}\\
   &=&\sum_{\bm\alpha} {\Gamma}^{(N)}_{\bm\alpha} C^{(N)\dag}_{\bm\alpha} |0\rangle\,,\label{st2}
   \end{eqnarray}
   \label{stN}
   \end{subequations}
$\!\!$where $\Gamma_{i_1,\ldots,i_N}$ is a fully symmetric (antisymmetric) tensor for bosons (fermions) and the sum over each $i_j$ in \eqref{st1} runs over all $d$ sp states, whereas that in \eqref{st2} over all distinct $d_N$ operators  \eqref{Cm}, with  (see App.\ \ref{A})
  \begin{equation}  {\Gamma}^{(N)}_{\bm\alpha}=\langle 0| C_{\bm\alpha}^{(N)}|\Psi\rangle=\frac{\Gamma_{i_1\ldots i_N}}{\sqrt{n_1!}\ldots\sqrt{n_d!}}\,,\label{gamnt}
  \end{equation}
   for     $c^\dag_{i_1}\ldots c^\dag_{i_N}=c_1^{\dag n_1}\ldots c_d^{\dag n_d}$ (and $i_1< \ldots<i_N$ for fermions). Here $|\Gamma_{\bm\alpha}^{(N)}|^2$ is the probability of finding the $N$ particle state $C^{(N)\dag}_{\bm{\alpha}}|0\rangle$ ``occupied'' in $|\Psi\rangle$, 
   with $\langle\Psi|\Psi\rangle=\sum_{\bm \alpha} |{\Gamma}^{(N)}_{\bm \alpha}|^2=\tfrac{1}{N!}\sum_{i_1,\ldots,i_d}|\Gamma_{i_1\ldots i_N}|^2=1$. 

\subsection{The $(M,N-M)$ representation and Schmidt decomposition for bosons and fermions}

We can rewrite the general $N$-particle state \eqref{stN} in a bipartite-like form involving operators creating  $M\leq N$ and  $N\!-\!M$ particles, such that side $A$  refers to $M$ particles (but not to any specific location in space or any other quantum number) and side $B$ to $N-M$ particles. Starting from Eq.\ \eqref{Nm} we obtain $\sum_{\bm\alpha} C^{(M)\dag}_{\bm\alpha}C_{\bm{\alpha}}^{(M)}|\Psi\rangle=\binom{N}{M}|\Psi\rangle$. Then  $|\Psi\rangle=
\binom{N}{M}^{-1} \sum_{\bm\alpha}C^{(M)\dag}_{\bm\alpha}C^{(M)}_{\bm\alpha}|\Psi\rangle$ can be  recast in the bipartite-like form 
\begin{eqnarray}|\Psi\rangle&=&
{\textstyle\binom{N}{M}^{-1}}\sum_{\bm\alpha, \bm\beta} {\Gamma}^{(M)}_{\bm\alpha\bm\beta}C^{(M)\dag}_{\bm\alpha} C^{(N-M)\dag}_{\bm\beta}|0\rangle\,,
   \label{Psm}
 \end{eqnarray}
for both bosons and fermions, where we have written 
\begin{eqnarray}C_{\bm\alpha}^{(M)}|\Psi\rangle
&=&\sum_{\bm\beta}{\Gamma}^{(M)}_{\bm\alpha\bm\beta}C^{(N-M)\dag}_{\bm\beta}|0\rangle\,,\label{Cbet}\label{9}
\end{eqnarray}
 and  sums over $\bm\alpha,\bm\beta$ run  over all $d_M$ and  $d_{N-M}$ operators $C^{(M)\dag}_{\bm\alpha}$,  $C^{(N-M)\dag}_{\bm\beta}$ 
respectively. Here 
$\Gamma^{(M)}_{\bm\alpha\bm\beta}\equiv\Gamma^{(M,N-M)}_{\bm\alpha\bm\beta}$ is given by (see Eq.\ \eqref{ort})
\begin{eqnarray}    {\Gamma}^{(M)}_{\bm\alpha\bm\beta}&=&\langle 0|C^{(N-M)}_{\bm\beta} C^{(M)}_{\bm\alpha}|\Psi\rangle\label{gdef1}
\,,
\end{eqnarray} 
and is directly related to $\Gamma_{i_1,\ldots,i_n}$ in  \eqref{st1} by Eq.\ \eqref{gammt}.     

Eq.\ \eqref{Psm} is the $(M,N\!-\!M)$ bipartite-like decomposition of $|\Psi\rangle$, expressing it as a linear combination of  ``products'' of  states in ${\cal F}_M$ and ${\cal F}_{N-M}$. The coefficients  ${\Gamma}^{(M)}_{\bm\alpha\bm\beta}$  determine the remnant  $N-M$ particle state \eqref{9} 
after annihilating in $|\Psi\rangle$ $M$ particles in the state labeled by ${\bm\alpha}$, with $|\Gamma^{(M)}_{\bm\alpha\bm\beta}|^2$ proportional to the  probability of having  $M$ particles in the state $\bm{\alpha}$ and $N-M$ in the state $\bm\beta$. 
 Eqs.\ \eqref{Psm} and \eqref{Cbet} imply 
$\langle\Psi|\Psi\rangle={\textstyle\binom{N}{M}^{-1}}\sum_{\bm\alpha,\bm\beta}|{\Gamma}^{(M)}_{\bm\alpha\bm\beta}|^2$, 
such that for any normalized state, 
\begin{equation}
{\rm Tr}\,[\Gamma^{(M)^\dag}\Gamma^{(M)}]=\binom{N}{M}
\label{tr1}\,,\end{equation}
for both bosons and fermions.

As done for fermions \cite{GDR.21},  
from the singular value decomposition (SVD)   of the matrix $\Gamma^{(M)}$,   \begin{equation}\Gamma^{(M)}_{\bm\alpha\bm\beta}=\sum_{\nu}U^{(M)}_{\bm\alpha\nu}\sigma_\nu^{(M)}V^{(N-M)\dag}_{\nu\bm\beta}
\label{SVD}
\,,\end{equation} 
where $U^{(M)},V^{(N-M)}$ are unitary $d_M\times d_M$ and    $d_{N-M}\times d_{N-M}$ matrices  and 
 $\sigma^{(M)}_\nu>0$
the singular values of $\Gamma^{(M)}$ 
(square roots of the nonzero eigenvalues of $\Gamma^{(M)\dag}\Gamma^{(M)}$ or $\Gamma^{(M)}\Gamma^{(M)\dag}$), we obtain from \eqref{Psm} {\it the Schmidt-like diagonal 
$(M,N\!-\!M)$ decomposition} of a general bosonic or fermionic $N$-particle state, 
\begin{equation}
    |\Psi\rangle={\textstyle\binom{N}{M}^{-1}}\sum_{\nu=1}^{n_{M}} \sigma^{(M)}_\nu A^{(M)\dag}_\nu B^{(N-M)\dag}_\nu|0\rangle\,,\label{ScDcx}
\end{equation}
where $n_M$ is the rank of $\Gamma^{(M)}$ and 
\begin{equation}    
\label{15}\begin{split}
A^{(M)\dag}_\nu&=\sum_{\bm\alpha}U^{(M)}_{\bm\alpha\nu}
    C^{(M)\dag}_{\bm\alpha}\,,\\
    B^{(N-M)\,\dag}_\nu&=\sum_{\bm\beta} V^{(N-M)*}_{\bm\beta\nu}C^{(N-M)\,\dag}_{\bm\beta}\,, 
\end{split}
\end{equation}
are ``collective'' operators creating $M$ and $N-M$  particles. As $U^{(M)}$ and $V^{(N-M)}$ are unitary, they  are again 
orthogonal normalized operators satisfying 
\begin{subequations}\label{16}\begin{eqnarray}A^{(M)}_\nu A^{(M)\dag}_{\nu'}|0\rangle&=\delta_{\nu\nu'}|0\rangle=&B^{(N-M)}_\nu B^{(N-M)\dag}_{\nu'}|0\rangle,\;\;\;\;\;\;\;\;\;\label{16a}\\
\sum_\nu A^{(M)\dag}_\nu A_\nu^{(M)}&=\;\;\binom{\hat{N}}{M}\;\;=&\sum_\nu B^{(N-M)\dag}_\nu B^{(N-M)}_\nu,\;\;\;\;\;\;\;\;\;\label{16b}\end{eqnarray}
\end{subequations}
for both bosons and fermions. 
Moreover,  Eqs.\ \eqref{Cbet}--\eqref{gdef1}  become  diagonal in terms of these normal operators:  
\begin{subequations}
\label{17}
\begin{eqnarray}
A_\nu^{(M)}|\Psi\rangle=\sigma^{(M)}_\nu B^{(N-M)\dag}_\nu|0\rangle\,,\label{17b}\\
\langle 0|B^{(N-M)}_{\nu}A_{\nu'}^{(M)}|\Psi\rangle=\delta_{\nu\nu'}\,\sigma_\nu^{(M)}\,,
\end{eqnarray}
\end{subequations}
such that $B^{(N-M)\dag}_\nu|0\rangle$ is the state of remaining $N-M$ particles after destroying $M$ particles in the state labeled by $\nu$. These states are orthogonal according to Eq.\ \eqref{16a}, in analogy with the standard Schmidt decomposition and in contrast with the states \eqref{Cbet} (see Eq.\ \eqref{M}).  On the other hand, the full terms 
$A^{(M)\dag}_\nu B^{(N-M)\dag}_\nu|0\rangle$ are not necessarily orthogonal  for different $\nu$. 

The singular values $\sigma^{(M)}_\nu$ in \eqref{ScDcx} are characteristic of the state, i.e.\ independent of the choice of sp basis used to represent it (see \eqref{Usp} in App.\ \ref{A}). 
 For $N=2$ Eq.\ \eqref{ScDcx} becomes equivalent to the normal  forms of refs.\ \cite{SC.01,ES.02}. 

 \subsection{The $M$-body density matrix and  operator}
The bipartite tensor $\Gamma^{(M)}_{\alpha\beta}$  is directly  connected to the {\it $M$-body density matrix} $\rho^{(M)}$,   of  elements \cite{RS.80,An.63}
\begin{subequations}
\label{18t}
\begin{eqnarray}
\rho^{(M)}_{\bm\alpha\bm\alpha'}&:=&\langle\Psi|C^{(M)\dag}_{\bm\alpha'}C^{(M)}_{\bm\alpha}|\Psi\rangle\,\label{M}\\&=&
\sum_{\bm\beta}\Gamma_{\bm\alpha\bm\beta}^{(M)}\Gamma_{\bm\alpha'\bm\beta}^{(M)*}=(\Gamma^{(M)}\Gamma^{(M)\dag})_{\bm\alpha\bm\alpha'}\,,\;\;\;\label{18}
\end{eqnarray}
\end{subequations}
i.e.\  $\rho^{(M)}=\Gamma^{(M)}\Gamma^{(M)\dag}$ for both bosons and fermions, where  in \eqref{18} we used Eqs.\ \eqref{ort}-\eqref{Cbet}.  
 Here $\rho^{(M)}$ is a $d_M\times d_M$ positive semidefinite matrix, 
 representing the hermitian {covariance matrix} 
 of the linearly independent operators $C^{(M)}_{\bm\alpha}$  in the  state $|\Psi\rangle$.  Eq.\   \eqref{tr1}  implies 
\begin{equation}
{\rm Tr}\,\rho^{(M)}=\binom{N}{M}\,,\label{tr2}
\end{equation}
for  both bosons and fermions, in agreement with Eq.\ \eqref{Nm}  \footnote{Here $\rho^{(M)}_{\bm\alpha\bm\alpha'}=\langle C^{(M)\dag}_{\bm\alpha'}C^{(M)}_{\bm\alpha}\rangle$ is defined over linearly independent operators $C_{\bm\alpha}^{(M)\dag}$ creating normalized  orthogonal $M$-particle states, 
such that its trace and eigenvalues differ by a factor $1/M!$ from those obtained with  definitions covering all averages  $\langle c^\dag_{i_1}\ldots c^\dag_{i_M}c_{i'_1}\ldots c_{i'_M}\rangle$ for arbitrary labels, 
where elements are then repeated $M!$ times. The rank of $\rho^{(M)}$ 
 remains nonetheless  unchanged $\forall M$.}.  The average of any bosonic or fermionic $M$-body operator can then be expressed as  
\begin{equation}
    \langle \sum_{\bm\alpha,\bm\alpha'}O^{(M)}_{\bm\alpha\bm\alpha'}C^{(M)\dag}_{\bm\alpha}C^{(M)}_{\bm\alpha'}\rangle= {\rm Tr}\,[\rho^{(M)}O^{(M)}]\,.\label{Om}
\end{equation}
Since Eq.\ \eqref{Psm} implies  $\Gamma^{(N-M)}=(\pm 1)^{M(N-M)}\Gamma^{(M)\,t}$ for bosons ($+$) or fermions ($-$), the partner DM $\rho^{(N-M)}_{\bm\beta\bm\beta'}=\Gamma^{(N-M)}\Gamma^{(N-M)\dag}$ 
 is just $\Gamma^{(M)t}\Gamma^{(M)*}$. 
 
From \eqref{SVD} we note that 
the squared singular values 
\begin{equation}
\lambda_\nu^{(M)}=(\sigma^{(M)}_\nu)^2\,,
\end{equation}
arising from the Schmidt decomposition  \eqref{ScDcx},  
are precisely the nonzero eigenvalues of $\Gamma^{(M)}\Gamma^{(M)^\dag}=\rho^{(M)}$ or equivalently $\Gamma^{(M) t}\Gamma^{(M)*}=\rho^{(N-M)}$, i.e., of the  $M$ and $N\!-\!M$-body DMs, which then have  the  same non-zero eigenvalues in any $N$-particle pure state $|\Psi\rangle$, for both  bosons and fermions \cite{CK.61},   
with $U^{(M)}$ and $V^{(N-M)*}$ the  corresponding eigenvector matrices. This result is analogous to  that of  the distinguishable bipartite case \cite{NC.00}. 

Moreover, the normal operators \eqref{15} are precisely those which {\it diagonalize} $\rho^{(M)}$ and $\rho^{(N-M)}$,  constituting the $M$ and $(N\!-\!M)$-body ``natural orbitals'':   
\begin{eqnarray}    \langle\Psi|A^{(M)\dag}_{\nu'}A^{(M)}_{\nu}|\Psi\rangle&=&
    \lambda_\nu^{(M)}\delta_{\nu\nu'}\\&=&
    \langle\Psi|B^{(N-M)\,\dag}_{\nu'} B^{(N-M)}_{\nu}|\Psi\rangle\,,\label{24}
    \nonumber
\end{eqnarray}
as follows from \eqref{16}--\eqref{17}. 

For $M=N$, $\rho^{(N)}$ has just a single eigenvalue $\lambda^{(N)}_1=1$ corresponding to the operator $A^{(N)\,\dag}_1=\sum_{\bm\alpha} \Gamma^{(N)}_{\bm\alpha} C^{(N)\dag}_{\bm\alpha}$ creating the  state \eqref{st2}, whereas for   $M=1$ we recover the {\it one-body} DM $\rho^{(1)}_{ij}=\langle\Psi|c^\dag _{j}c_{i}|\Psi\rangle=(\Gamma^{(1)}\Gamma^{(1)^\dag})_{ij}$, with ${\rm Tr}\,\rho^{(1)}=N$, isospectral with $\rho^{(N-1)}$ \cite{GDR.20}. In this case $A^\dag_\nu=c^\dag_\nu$ are the standard sp ``natural'' creation operators diagonalizing $\rho^{(1)}$: $\langle c^\dag_{\nu'}c_\nu\rangle=\delta_{\nu\nu'}\lambda_\nu^{(1)}$. 

We also mention that in an $N$-particle Fock state $|\Psi_{\bm\beta}\rangle=C^{(N)\dag}_{\bm\beta}|0\rangle=\frac{c_1^{\dag n_1}}{\sqrt{n_1!}}\ldots\frac{c_d^{\dag n_d}}{\sqrt{n_d}}|0\rangle$ ($\bm\beta=(n_1,\ldots,n_d)$, $\sum_i n_i=N$), i.e.\ a permanent (bosons) or SD (fermions), $\rho^{(M)}$ is diagonal in the standard basis of operators $C^{(M)\dag}_{\bm\alpha}=\frac{c_1^{\dag m_1}}{\sqrt{m_1!}}\ldots\frac{c_d^{\dag m_d}}{\sqrt{m_d!}}$ 
($\bm\alpha=(m_1,\ldots,m_d)$, $\sum_i m_i=M$), having just integer eigenvalues: 
\begin{equation}\langle C^{(M)\dag}_{\bm\alpha}C^{(M)}_{\bm\alpha'}\rangle_{\bm\beta}=\delta_{\bm\alpha\bm\alpha'}\lambda^{(M)}_{\bm\alpha}\,,\;\;\;\lambda_{\bm\alpha}^{(M)}={\textstyle\prod_i\binom{n_i}{m_i}}\,,\end{equation}
which in the fermionic case reduce just to $\binom{N}{M}$  eigenvalues $\lambda^{(M)}_{\bm\alpha}=1$ \cite{GDR.21}. In both cases they  verify  $\sum_{\bm\alpha}\lambda_{\bm\alpha}^{(M)}=\binom{N}{M}$. In the bosonic case the lowest rank obviously corresponds to a condensate (e.g.\ $n_i=N\delta_{i1}$) where there is a single nonzero eigenvalue $\lambda^{(M)}_{1}=\binom{N}{M}$.  

We can also define the {\it $M$-body density operator (DO)} 
\begin{subequations}
\begin{eqnarray} \hat\rho^{(M)}&=&\sum_{\bm \beta}C^{(N-M)}_{\bm\beta}|\Psi\rangle\langle\Psi|C_{\bm \beta}^{(N-M)\dag}\label{r1o}\\
&=&\sum_{\bm \alpha,\bm\alpha'} \rho^{(M)}_{\bm\alpha\bm\alpha'}C^{(M)\dag}_{\bm\alpha}|0\rangle\langle 0|C^{(M)}_{\bm\alpha'}\,,\label{r2o}\end{eqnarray}
\label{rhomo}
\end{subequations}
where in \eqref{r2o} we used  Eq.\ \eqref{Cbet} for $M\rightarrow N\!-\!M$ and  \eqref{18}. 
It is the unique mixed $M$-particle state  fulfilling   
\begin{equation}{\rm Tr}\,[\hat\rho^{(M)}C^{(M)\dag}_{\bm\alpha'}C^{(M)}_{\bm\alpha}]=\rho^{(M)}_{\bm\alpha\bm\alpha'}
\end{equation}
$\forall\,\bm\alpha,\bm\alpha'$, for both  bosons or fermions. 
  Its diagonal form is  \begin{eqnarray} \hat\rho^{(M)}&=&\sum_\nu\lambda_\nu^{(M)}A^{(M)\dag}_\nu|0\rangle\langle 0|A_\nu^{(M)}\,,\end{eqnarray}
in terms of the normal operators \eqref{15}, such that  $\hat\rho^{(M)}A^{(M)\dag}_\nu|0\rangle=\lambda_\nu^{(M)}A^{(M)\dag}_\nu|0\rangle$, 
having obviously the same eigenvalues as its matrix representation $\rho^{(M)}$. For $M=N$ it reduces to $\hat\rho^{(N)}=|\Psi\rangle\langle\Psi|$. 

Thus, the operation in \eqref{r1o} can be seen as a partial trace over $N-M$ particles,  leading to the reduced state $\hat\rho^{(M)}$ of Eq.\ \eqref{r2o} which determines the average of any $M$-body (and hence $L$-body for $L<M$, see \eqref{27}) operator, in analogy with the standard distinguishable case.

Under unitary sp transformations of the state, $|\Psi\rangle\rightarrow \hat U|\Psi\rangle$ with $\hat U=e^{-\imath\sum_{ij} h_{ij}c^\dag_i c_j}$,  all $\rho^{(M)}$ and $\hat\rho^{(M)}$  will  transform unitarily (see \eqref{Usp2}). \\

\subsection{Measurements,   $M$-body DM's  and  $M$-body entanglement}
For both bosons and fermions, the operators 
\begin{equation} {\cal M}_{\bm\beta}:= {\textstyle\binom{N}{M}^{-1/2}} C^{(N-M)}_{\bm\beta}\,,\label{Kr}\end{equation}
can be considered as Kraus operators when acting on the subspace of states with definite particle number $N$, since by  Eq.\ \eqref{Nm},  they satisfy   
\begin{equation}\sum_{\bm\beta}\,{\cal M}^\dag_{\bm\beta} {\cal M}_{\bm\beta}=\mathbbm{1}_N\,,\label{Kr2}\end{equation}
i.e.\ $\sum_{\bm\beta}{\cal M}^\dag_{\bm\beta} {\cal M}_{\bm\beta}|\Psi\rangle=|\Psi\rangle$ $\forall$ $N$-particle state $|\Psi\rangle$. 
Then they define a measurement on $N$-particle states in which  $N-M$ particles are annihilated (see also \eqref{B7} in App.\ \ref{B} for a number conserving implementation): $|\Psi\rangle\rightarrow {\cal M}_{\bm\beta}|\Psi\rangle\propto C^{(N-M)}_{\bm\beta}|\Psi\rangle$, with probabilities 
\begin{equation}
p_{\bm\beta}=\langle \Psi|{\cal M}^{\dag}_{\bm\beta}{\cal M}_{\bm\beta}|\Psi\rangle=\rho^{(N-M)}_{\bm\beta\bm\beta}/{\textstyle\binom{N}{M}}\,,\label{pbet}\end{equation} 
 determined precisely by the $(N-M)$-body DM. 

And according to Eqs.\ \eqref{r1o}-\eqref{Kr},  the ensuing post-measurement state $\sum_\beta {\cal M}_{\bm \beta}|\Psi\rangle\langle\Psi|{\cal M}_{\bm \beta}^\dag$  (without postselection) is just the 
{\it normalized $M$-body DO} $\hat\rho_n^{(M)}$:
\begin{eqnarray}
\hat\rho^{(M)}_n&:=&\!{\textstyle\binom{N}{M}^{-1}}\!\hat\rho^{(M)}=
\sum_{\bm\beta}\! {\cal M}_{\bm\beta}|\Psi\rangle\langle\Psi| {\cal M}^{\dag}_{\bm\beta}\,,
\label{31}
\end{eqnarray}
which satisfies ${\rm Tr}\,\hat\rho^{(M)}_n=1$.  The basic case is that of sp measurements ($N-M = 1$, ${\cal M}_{i}=c_i/\sqrt{N}$),  with $N-M$-body measurements based on the operators \eqref{Kr} being just compositions of sp measurements.

Regarding  the  $L$-body DM $\rho^{(L)}$ for $L\leq M$,  we first note that using   Eqs.\ \eqref{r1o} and \eqref{Nm}, it can be obtained from $\hat\rho^{(M)}$ as (see App.\ \ref{A}) 
\begin{equation}\rho^{(L)}_{\bm\gamma\bm\gamma'}= {\textstyle\binom{N-L}{N-M}}^{-1}{\rm Tr}\,[\hat\rho^{(M)}C^{(L)\dag}_{\bm\gamma'}C^{(L)}_{\bm\gamma}]
\;\;\;(L\leq M).\label{27}\end{equation} 
 Then the expansions \eqref{r1o} or \eqref{31} imply  (see App.\ \ref{A}) 
 \begin{equation} \hat{\rho}_n^{(L)}=\sum_{\bm \beta}p_{\bm\beta}\,\hat\rho^{(L)}_{\bm \beta n}\,,\label{res}\end{equation}
  where $\hat\rho^{(L)}_{n}=\hat\rho^{(L)}/\binom{N}{L}$ is the normalized $L$-body DO in the original state $|\Psi\rangle$ and $\hat\rho^{(L)}_{\bm\beta n}=\hat\rho^{(L)}_{\bm\beta}/\binom{M}{L}$ those in the postselected normalized $M$-particle states $|\Psi_{\bm\beta}\rangle={\cal M}_{\bm\beta}|\Psi\rangle/\sqrt{p_{\bm\beta}}$, with $p_{\bm\beta}$ the probabilities \eqref{pbet} 
   of outcome $\bm{\beta}$.  
    Thus, for any $L\leq M$, 
  the normalized DM $\rho^{(L)}_n$ is the average  of the normalized post-measurement DMs $\hat\rho^{(L)}_{\bm\beta n}$ in the post selected states. 
  
  This result  is important since it implies the general majorization relation \cite{GDR.21}
\begin{equation} \bm{\lambda}(\rho_n^{(L)})\prec \sum_{\bm\beta}p_{\bm\beta}\bm{\lambda}(\hat\rho^{(L)}_{\bm\beta n})\label{prec}\end{equation}
between the sorted (in decreasing order) eigenvalue spectrum $\bm\lambda$ of the original DM $\rho_n^{(L)}$ and those of $\hat\rho_{\bm\beta n}^{(L)}$ (see e.g.\ \cite{Bh.97,Moa.11} for  majorization properties). It  entails the general entropic inequality \begin{equation}S(\hat\rho_n^{(L)})\geq \sum_{\bm\beta}p_{\bm\beta}S(\hat\rho^{(L)}_{\bm\beta n})\label{sentt}\,,
\end{equation}
between the entropy of the original normalized $L$-body DM $\rho_n^{(L)}$  and the average entropy of the normalized $L$-body DMs in the post-measurement states. It is valid for {\it any} concave entropy   $S(\rho)$,  like the von Neumann entropy $S(\rho)=-{\rm Tr}\,\rho\log_2\rho$,  or in general any trace-form entropy $S_f(\rho)={\rm Tr}\,f(\rho)$ with $f$ concave and $f(0)=f(1)=0$ \cite{RC.022}, 
in both boson and fermion systems. 

This result means that for $L\leq M$, the {\it $L$-body entanglement}, determined by the mixedness of the normalized $L$-body DM \cite{GDR.21} and quantified by the associated entropy 
    \begin{equation} E^{(L)}(|\Psi\rangle)=S(
\hat\rho^{(L)}_n)=S(
\hat\rho^{(N-L)}_n)\label{EntL}\,,\end{equation}  
 {\it cannot  increase} (and will typically decrease) on average under the  $N\!-\!M$-body operation determined by the operators \eqref{Kr}, for both bosons  and fermions,  and for {\it any} choice of entropy.  This  is in agreement with the fact that such measurement decreases the uncertainty about which $L$-body states are occupied. 
  
  Eqs.\ \eqref{Kr2}--\eqref{EntL} remain  valid for measurements based on the normal operators \eqref{15},  $\tilde{\cal M}_\nu=\binom{N}{M}^{-\frac{1}{2}}B^{(N-M)}_\nu$, 
  as they are unitarily related to the ${\cal M}_{\bm \beta}$:  
  just replace $\bm\beta\rightarrow \nu$, with $p_{\bm\beta}\rightarrow p_\nu=\lambda_\nu^{(N-M)}/{\textstyle\binom{N}{M}}$ 
 and $|\Psi_{\bm\beta}\rangle\rightarrow |\Psi_\nu\rangle=A_{\nu}^{(M)\dag}|0\rangle$ ($\propto B^{(N-M)}_\nu|\Psi\rangle$)   in \eqref{res}--\eqref{sentt}. 

Hence, a  resource theory based on the previous quantum operations  (plus  free operations like unitary sp transformations) is in principle feasible, with the $L$-body entanglement,  determined by the mixedness of the reduced densities  $\hat\rho^{(L)}_n$ (which is fully independent of the choice of sp modes) as a basic resource which cannot  increase on average under these operations (see also App.\ \ref{B}). 

\section{Reduced exact decompositions\label{III}}

The presence of symmetries in $|\Psi\rangle$ can simplify  the DMs $\rho^{(M)}$ into a blocked structure in an obvious basis, reducing the effective number of nonzero elements. A common example is 
 the conservation of the number of particles in a certain subspace $\cal{S}\subset{\cal H}$ of sp modes, \begin{equation}\hat{N}_{\cal S}=\sum_{i\in {\cal S}}c^\dag_i c_i\label{Ns0}\,,\end{equation}   as in the case of  eigenstates $|\Psi\rangle$ of Hamiltonians satisfying $[H,\hat{N}_{\cal S}]=0$, such that 
 \begin{equation}
    \hat{N}_{\cal S}|\Psi\rangle=N_{\cal S}|\Psi\rangle\,. 
   \label{Ns}\end{equation} 
   If $|\Psi\rangle$ has definite  particle number $N$,    
\eqref{Ns} also implies $\hat{N}_{\bar{{\cal S}}}|\Psi\rangle=N_{\bar{{\cal S}}}|\Psi\rangle$ for $N_{\bar{{\cal S}}}=N\!-\!N_{\cal S}$ the number of particles in the orthogonal complement $\bar{\cal S}$, such that ${\cal H}={\cal S}\oplus\bar{\cal S}$. 

 Common well-known cases are e.g.\ systems with pairing-type two-body couplings,  where the number of particles in positive and negative quasimomentum states $N_{\cal S}$, $N_{\cal \bar S}$, are  conserved  (see sec.\  \ref{IV}), 
  Hubbard-type Hamiltonians in solid state physics 
 \cite{Fe.22,Hub.63,PRX15H}, which conserve the number of particles $N_\sigma=\sum_{i}c^\dag_{i\sigma}c_{i\sigma}$ with definite spin component $\sigma$, and the strong nuclear force in an atomic nucleus within the isospin formalism \cite{RS.80}, which conserves $T_z=\frac{N-Z}{2}$, i.e., the number of  neutrons $N=\sum_k c^\dag_{k+}c_{k+}$ and protons $Z=\sum_k c^\dag_{k-}c_{k-}$ for $N+Z$ fixed and $k$ denoting remaining quantum numbers. It is also the case of any system with fixed number of particles at distinct sites (corresponding to orthogonal sp subspaces ${\cal S}_i$), like entangled states of  spatially separated $M$ and $N-M$ particles, where the standard distinguishable scenario emerges naturally as a special case (sec.\ \ref{IIID}).  
   
Eq.\ \eqref{Ns} implies that elements of $\rho^{(M)}$ which do not conserve the number of particles in ${\cal S}$ will vanish: $\langle C^{(M)\dag}_{\bm\alpha}C^{(M)}_{\bm\alpha'}\rangle=0$ if $[ C^{(M)\dag}_{\bm\alpha}C^{(M)}_{\bm\alpha'},\hat{N}_{\cal S}]\neq 0$, leading to a blocked $\rho^{(M)}$ where each block corresponds to  a fixed  number  $m$ of operators $c^\dag_{i\in {\cal S}},c_{i'\in {\cal S}} $ in  $C^{(M)\dag}_{\bm\alpha}$,  $C^{(M)}_{\bm\alpha'}$.      
 We show here that reduced exact $(M,N\!-\!M)$ expansions of  $|\Psi\rangle$  associated with each of these blocks are also  feasible. 
 
\subsection{One-body case}
We start with the simplest case $M=1$. It is easily seen that Eq.\ \eqref{Ns} implies the  following blocked form of
 the one-body DM $\rho^{(1)}$, 
\begin{equation}
    \rho^{(1)}=\begin{pmatrix}
    \rho^{(1)}_{\cal S}&0\\0&\rho^{(1)}_{\bar{{\cal S}}}
    \end{pmatrix}\,,\label{r1}
\end{equation}
where $(\rho^{(1)}_{{\cal S}})_{ij}=\langle c^\dag_j c_i\rangle$, $(\rho^{(1)}_{\bar{{\cal S}}})_{ij}=\langle c^\dag_{\bar{j}} c_{\bar{i}}\rangle$ are the one-body DM's in each subspace, since remaining contractions $\langle c^\dag_ic_{\bar{j}}\rangle$ vanish due to  the conservation of $\hat{N}_{\cal S}$.  
 These blocks have a fixed trace ${\rm Tr}\,\rho^{(1)}_{{\cal S}}=N_{\cal S}$, ${\rm Tr}\,\rho^{(1)}_{\bar{{\cal S}}}=N_{\bar{{\cal S}}}$.  

Moreover, if $N_{\cal S}\geq 1$, $N_{\bar{{\cal S}}}\geq 1$, each block can  be associated to an own  $(1,N\!-\!1)$ expansion and Schmidt decomposition of $|\Psi\rangle$:  Starting from \eqref{Ns}, $|\Psi\rangle=\frac{1}{N_{\cal S}}\hat{N}_{\cal S}|\Psi\rangle$, 
 and writing $c_i|\Psi\rangle=\sum_{\bm\beta}
 \Gamma^{(1)}_{i\bm\beta}C^{(N-1)\dag}_{\bm\beta}|0\rangle$ with $\Gamma^{(1)}_{i\bm\beta}=\langle 0|C_{\bm\beta}^{(N-1)}c_i|\Psi\rangle$, we obtain the reduced exact expansion  
\begin{subequations}
\begin{eqnarray}
|\Psi\rangle&=&
N_{\cal S}^{-1}
\!\sum_{i\in {\cal S},\bm\beta}\Gamma^{(1)}_{i\bm\beta}c^\dag_iC^{(N-1)\dag}_{\bm\beta}|0\rangle\\
&=&
N_{\cal S}^{-1}\sum_{\nu\in {\cal S}}\!\sigma^{(1)}_{\nu} a^\dag_{\nu}B^{(N-1)\dag}_\nu|0\rangle\label{40b}\,,
\end{eqnarray}
\label{rexp}
\end{subequations}
$\!$which just involves the  block $\Gamma^{(1)}_{\cal S}$ of  elements $\Gamma^{(1)}_{i\in {\cal S},\bm\beta}$ of the full  $\Gamma^{(1)}$ in \eqref{Psm}, where $\bm\beta$ spans $(N\!-\!1)$-particle states with $N_{\cal S}\!-\!1$ particles  in ${\cal S}$ and $N_{\bar{{\cal S}}}$ in $\bar{{\cal S}}$. In \eqref{40b}, $\sigma_{\nu}^{(1)}$ are the singular values of $\Gamma^{(1)}_{\cal S}$ and $a_\nu=\sum_{i\in {\cal S}}U^{(1)}_{i\nu}c^\dag_i$, $B^\dag_\nu=\sum_{\bm\beta}V^{(N-1)*}_{\bm\beta\nu} C^{(N-1)\dag}_{\bm\beta}$  the associated normal operators \eqref{15} 
($M=1$).  It leads to $\langle c^\dag_j c_i\rangle=(\Gamma^{(1)}_{\cal S}\Gamma^{(1)\dag}_{\cal S})_{ij}$ for $i,j\in {\cal S}$ and hence to the upper block of $\rho^{(1)}$, 
\begin{equation}
    \rho_{\cal S}^{(1)}=\Gamma_{\cal S}^{(1)}\Gamma^{(1)\dag}_{\cal S}\,,\label{rho1s}
\end{equation}
with eigenvalues  $\lambda^{(1)}_{\cal \nu}=(\sigma_{\nu}^{(1)})^2$ for $\nu\in{\cal S}$. 

Similar expressions with 
${\cal S}\rightarrow \bar{\cal S}$
in \eqref{rexp}-\eqref{rho1s} obviously hold for the expansion of $|\Psi\rangle$ associated to  the second block  $\rho^{(1)}_{\bar{{\cal S}}}$ in \eqref{r1}, determined by $\Gamma^{(1)}_{\bar{{\cal S}}}$  of  elements $\Gamma^{(1)}_{\bar{i}\in\bar{{\cal S}},\bm\beta}$.  Both expansions, Eq.\  \eqref{rexp} and the analogous one based on $\bar{\cal S}$, are exact but run in general over distinct singular values $\sigma_\nu^{(1)}$, $\sigma_{\bar\nu}^{(1)}$ and  operators $a_\nu$, $a_{\bar\nu}$. For composite systems with $N_{{\cal S}_i}\geq 1$ particles in $n$ orthogonal  subspaces ${\cal S}_i$, analogous  expansions hold for each subspace.

Eq.\ \eqref{Ns} also leads to a similar blocked structure of the partner isospectral $(N\!-\!1)$-body DM, \begin{equation}
    \rho^{(N-1)}=\begin{pmatrix}
    \rho^{(N-1)}_{\cal S}&0\\0&\rho^{(N-1)}_{\bar{{\cal S}}}
    \end{pmatrix}
\end{equation}
where 
$\rho^{(N-1)}_{{\cal S}}=\Gamma_{\cal S}^{(1)T}\Gamma^{(1)*}_{\cal S}$ contains just those elements 
$\rho^{(N-1)}_{\bm\beta'\bm\beta}$ with $\bm\beta,\bm\beta'$ involving $N_{\cal S}-1$ particles in ${\cal S}$, and $\rho^{(N-1)}_{\bar{{\cal S}}}=\Gamma_{\bar{{\cal S}}}^{(1)T}\Gamma^{(1)*}_{\bar{{\cal S}}}$ those with $N_{\bar{\cal S}}-1$ particles in $\bar{\cal S}$ (and hence $N_{\cal S}$ in ${\cal S}$). 
 Notice, however,  that  $\rho^{(1)}_{\cal S}$ and $\rho^{(1)}_{\bar{{\cal S}}}$ are not isospectral in general. 
 
\subsection{Two-body case}
 The implications of \eqref{Ns} become even more important  for the two-body DM. 
For  $N_{\cal S}\geq 2$,  $N_{\bar{{\cal S}}}\geq 2$, Eq.\ \eqref{Ns} entails that the full $\rho^{(2)}$ will have  three principal blocks: 
\begin{equation}
    \rho^{(2)}=\begin{pmatrix}
    \rho^{(2)}_{\cal S}&0&0\\0&\rho^{(2)}_{{\cal S}\bar{\cal S}}&0\\
    0&0&\rho^{(2)}_{\bar{\cal S}}
    \end{pmatrix}\,,\label{rho2ss}
\end{equation}
where  
\begin{eqnarray}
(\rho^{(2)}_{\cal S})_{ij,kl}=\langle c^\dag_kc^\dag_l c_j c_i\rangle,&&
(\rho^{(2)}_{\bar{\cal S}})_{\bar i\bar j,\bar k\bar l}=\langle c^\dag_{\bar k}c^\dag_{\bar l} c_{\bar j} c_{\bar i}\rangle,\;\;\;
\end{eqnarray}
contain the contractions within ${\cal S}$ and $\bar{\cal S}$ respectively (including  diagonal elements $(\rho_{\cal S}^{(2)})_{ii,kl}=\langle c^\dag_k c^\dag_l c_i^2\rangle/\sqrt{2}$, 
$(\rho_{\cal S}^{(2)})_{ii,kk}=\langle c^{\dag 2}_{k}c_{i}^2\rangle/2$, etc.\  in the bosonic case) and  \begin{eqnarray}
(\rho^{(2)}_{{\cal S}\bar{{\cal S}}})_{i\bar j,k \bar l}
&=&\langle c^\dag_kc^\dag_{\bar l} c_{\bar j} c_{i}\rangle,
\label{reduced2}
\end{eqnarray}
  those  involving one particle in ${\cal S}$ and one in $\bar{\cal S}$.  All remaining contractions vanish due to the conserved  $N_{\cal S}$.  These blocks
have fixed traces 
\begin{equation}
{\rm Tr}\,\rho^{(2)}_{\cal S}={\textstyle\binom{N_{\cal S}}{2},\;
{\rm Tr}\,\rho^{(2)}_{\bar{{\cal S}}}={\textstyle\binom{N_{\bar{\cal S}}}{2}},\;
{\rm Tr}\,\rho^{(2)}_{{\cal S}\bar{\cal S}}=N_{\cal S} N_{\bar{{\cal S}}}},\label{tr2ssb}
\end{equation}
verifying $\binom{N_{\cal S}}{2}+\binom{N_{\bar{{\cal S}}}}{2}+N_{\cal S} N_{\bar{\cal S}}=\binom{N_{\cal S}+N_{\bar{\cal S}}}{2}$. 

Moreover,  Eq.\ \eqref{Ns} implies 
\begin{equation}\sum_{\bm\alpha\in{\cal S}}C^{(2)\dag}_{\bm\alpha}
C^{(2)}_{\bm\alpha}|\Psi\rangle={\textstyle\binom{N_{\cal S}}{2}}|\Psi\rangle\,,\label{sa1}\end{equation}
 for 
$C^{(2)\dag}_{\bm\alpha\in{\cal S}}\equiv c^\dag_i c^\dag_{j}$ ($i<j$) or $(c^\dag_i)^2/2$ , with a similar expression for ${\cal S}\rightarrow\bar{\cal S}$, and also 
\begin{equation}\sum_{\bm\gamma\in{\cal S}\bar{\cal  S}}C^{(2)\dag}_{\bm\gamma}
C^{(2)}_{\bm\gamma}|\Psi\rangle=N_{\cal S}N_{\bar{\cal S}}|\Psi\rangle\,.\label{sa2}
\end{equation}
for $C^{(2)\dag}_{\bm\gamma\in{\cal S}\bar{\cal S}}\equiv c^\dag_i c^\dag_{\bar j}$.  
Then the tensor $\Gamma^{(2)}$ will also have three 
corresponding blocks $\Gamma^{(2)}_{{\cal S}}$, $\Gamma^{(2)}_{{\cal S}\bar{{\cal S}}}$ and $\Gamma^{(2)}_{\bar{{\cal S}}}$, 
 each of them generating an {\it exact} reduced expansion and Schmidt decomposition of $|\Psi\rangle$:    The first one, emerging from \eqref{sa1}, 
\begin{subequations}\begin{eqnarray}
|\Psi\rangle&=&{\textstyle\binom{N_{\cal S}}{2}}^{-1}\!\!
\sum_{\bm\alpha\in S,\bm\beta}(\Gamma^{(2)}_{\cal S})_{\bm\alpha\bm\beta}C^{(2)\dag}_{\bm\alpha}C^{(N-2)\dag}_{\bm\beta}|0\rangle\;\;\\
&=&{\textstyle\binom{N_{\cal S}}{2}^{-1}}\sum_{\nu\in{\cal S}}\sigma_\nu^{(2)} A^{(2)\dag}_{\nu}B^{(N-2)\dag}_\nu|0\rangle\,,\end{eqnarray}\label{e1}\end{subequations}
where $(\Gamma^{(2)}_{\cal S})_{\bm\alpha\bm\beta}=\langle 0|
C_{\bm \beta}^{(N-2)}C_{\bm\alpha\in {\cal S}}^{(2)}|\Psi\rangle$  and $\sigma^{(2)}_\nu$ are the singular values of $\Gamma^{(2)}_{\cal S}$, is related to  $\rho^{(2)}_{\cal S}=\Gamma^{(2)}_{\cal S}\Gamma^{(2)\dag}_{\cal S}$ and
 its eigenvalues $\lambda^{(2)}_{\nu}=(\sigma_\nu^{(2)})^2$ for $\nu\in{\cal S}$.  The third one is analogous for  ${\cal S}\rightarrow \bar{\cal S}$ and is related to $\rho^{(2)}_{\bar{\cal S}}=\Gamma^{(2)}_{\bar{\cal S}}\Gamma^{(2)\dag}_{\bar{\cal S}}$. 
 
   Finally,   the second one, emerging from \eqref{sa2},  
\begin{subequations}\label{e3}\begin{eqnarray}
|\Psi\rangle&=&{\textstyle\frac{1}{N_{\cal S} N_{\bar{\cal S}}}}\!\!
\sum_{\bm\gamma
\in {\cal S}\bar{\cal S},\bm{\beta}}(\Gamma^{(2)}_{{\cal S}\bar{\cal S}})_{\bm\gamma\bm\beta}C^{(2)\dag}_{\bm\gamma}C^{(N-2)\dag}_{\bm\beta}|0\rangle\;\;\\
&=&{\textstyle\frac{1}{N_{\cal S}N_{\bar{\cal S}}}}\sum_{\tilde\nu}\sigma^{(2)}_{\tilde\nu}A^{(2)\dag}_{\tilde\nu}B^{(N-2)\dag}_{\tilde\nu}|0\rangle\label{e3b}\,,
\end{eqnarray}
\end{subequations}
where   $(\Gamma^{(2)}_{{\cal S}\bar{\cal S}})_{\bm\gamma\bm\beta}=\langle 0|
C_{\bm \beta}^{(N-2)}C_{\bm\gamma\in{\cal S}\bar{\cal S}}^{(2)}|\Psi\rangle$ and $\sigma^{(2)}_{\tilde\nu}$ are the singular values of $\Gamma^{(2)}_{{\cal S}\bar{\cal S}}$, 
determines the central block $\rho^{(2)}_{{\cal S}\bar{\cal S}}=\Gamma^{(2)}_{{\cal S}\bar{\cal S}}\Gamma^{(2)\dag}_{{\cal S}\bar{\cal S}}$ and its eigenvalues $\lambda^{(2)}_{\tilde\nu}=(\sigma_{\tilde\nu}^{(2)})^2$. It exposes the two-body  correlations between the particles in ${\cal S}$ and those in $\bar{\cal S}$. 
Summing the three previous expansions with their relative weights $p_{\cal S}=\binom{N_{\cal S}}{2}/\binom{N}{2}$, $p_{\bar{\cal S}}=\binom{N_{\bar{\cal S}}}{2}/\binom{N}{2}$ 
 and $p_{\cal S\bar{\cal S}}=N_{\cal S}N_{\bar{\cal S}}/\binom{N}{2}$ leads to the original expansion \eqref{Psm} for $M=2$.  
  
 Previous blocked structure and expansions also hold for the partner $(N\!-\!2)$-body DM $\rho^{(N-2)}$, with $\rho^{(N-2)}_{\cal S}=\Gamma^{(2)T}_{\cal S}\Gamma^{(2)*}_{\cal S}$, 
 $\rho^{(N-2)}_{{\cal S}\bar{\cal S}}=\Gamma^{(2)T}_{\cal S\bar{\cal S}}\Gamma^{(2)*}_{{\cal S}\bar{\cal S}}$ and $\rho^{(N-2)}_{\bar{\cal S}}=\Gamma^{(2)T}_{\bar{\cal S}}\Gamma^{(2)*}_{\bar{\cal S}}$ containing elements involving  $N_{\cal S}-2$, $N_{\cal S}-1$ and $N_{\cal S}$ particles in ${\cal S}$ respectively. 
 
The expansion \eqref{e3} is convenient when $\rho^{(2)}_{{\cal S}\bar{\cal S}}$ possesses one or a few large dominant eigenvalues $\lambda^{(2)}_{\tilde\nu}>1$ which absorb most of the sum  (see sec.\ \ref{IV}), and   which reflect pairing-like correlations  between particles  in  ${\cal S}$ and those in $\bar{\cal S}$ \cite{Samb.20,Yang.62,KW.09}. For any ``product'' state  \begin{subequations}\begin{eqnarray}|\Psi\rangle&=&A^{(N_{\cal S})\dag}_{\cal S}B^{(N_{\bar{\cal S}})\dag}_{\bar{\cal S}}|0\rangle\,,\label{psiprod}\end{eqnarray}  
where $A^{(N_{\cal S})\dag}_{\cal S}$ ($B^{(N_{\bar{\cal S}})\dag}_{\bar{\cal S}}$) creates an arbitrary state of $N_{\cal S}$  ($N_{\bar{\cal S}}$) particles in ${\cal S}$ ($\bar{\cal S}$), $\rho^{(2)}_{{\cal S}\bar{\cal S}}$  becomes a direct product of one-body densities in both fermion and boson systems,  
\begin{eqnarray}
(\rho^{(2)}_{{\cal S}\bar{\cal S}})_{i\bar{j},k\bar{l}}=
(\rho^{(1)}_{\cal S})_{ik}(\rho^{(1)}_{\bar{\cal S}})_{\bar{j}\bar{l}}\,,\label{prod1}
  \end{eqnarray}
  \end{subequations}
 becoming diagonal in the natural sp bases which diagonalize $\rho^{(1)}_{\cal S}$ and $\rho^{(1)}_{\bar{\cal S}}$:      $(\rho^{(2)}_{{\cal S}\bar{\cal S}})_{i\bar{j},k\bar{l}}=\delta_{ik}\delta_{\bar{j}\bar{l}}\lambda^{(1)}_{i}\lambda^{(1)}_{\bar{j}}$, with $\lambda^{(1)}_{i}\lambda^{(1)}_{\bar{j}}\leq 1$ for fermions. 

Hence an immediate   consequence for fermions is the following:
{ If 
in a state with 
definite fermion numbers $N_{\cal S}$, $N_{\bar{\cal S}}$ in  orthogonal sp subspaces ${\cal S}$, $\bar{\cal S}$, 
the joint two-body DM  $\rho^{(2)}_{{\cal S}\bar{\cal S}}$  has an eigenvalue $\lambda^{(2)}_\nu>1$, there is  {\it bipartite entanglement between the $N_{\cal S}$ and $N_{\bar{\cal S}}$ fermions}, in the sense of not being  a product state \eqref{psiprod} (see   \ref{IIID}).

Such dominant eigenvalue  also indicates  that $|\Psi\rangle$ cannot be an independent fermion state (SD) either, since in these states all nonzero eigenvalues of $\rho^{(2)}$ have the value $1$. Moreover, if $|\Psi\rangle$ is a SD ($(\rho^{(1)})^2=\rho^{(1)}$) and $N_{\cal S}$,  $N_{\bar{\cal S}}$ are conserved, then \eqref{psiprod} necessarily holds, 
with $A^{(N_{\cal S})\dag}_{\cal S}=C^{(N_{\cal S})\dag}_{\bm\alpha\in{\cal S}}$, $B^{(N_{\bar{\cal S}})\dag}_{\bar{\cal S}}=C^{(N_{\bar{\cal S}})\dag}_{\bm\beta\in\bar{\cal S}}$ simple product operators of the form \eqref{Cm}, since the blocked structure \eqref{r1} then implies $(\rho^{(1)}_{\cal S})^2=\rho^{(1)}_{\cal S}$, $(\rho^{(1)}_{\bar{\cal S}})^2=\rho^{(1)}_{\bar{\cal S}}$, entailing that both $A^{(N_{\cal S})\dag}_{\cal S}|0\rangle$, 
$B^{(N_{\bar\cal S})\dag}_{\cal S}|0\rangle$ are SDs. 

\subsection{General M-body case}
For a  general $M$-body DM in a system with $N_{\cal S}\geq M$ particles in  ${\cal S}$ 
and $N_{\bar{\cal S}}\geq M$ in $\bar{\cal S}$,  
Eq.\ \eqref{Ns} implies that 
$\rho^{(M)}$ will be blocked into $M+1$ subdensities   $\rho^{(m,l)}_{{\cal S}\bar{\cal S}}$ involving $m$ particles in ${\cal S}$ and $l=M-m$ particles in $\bar{\cal S}$, with $m=0,1,\ldots,M$:  
\begin{eqnarray}
\rho^{(M)}&=&\begin{pmatrix}\rho^{(M,0)}_{\cal S}&0&\ldots&
0\\0&\rho^{(1,M-1)}_{{\cal S}\bar{\cal S}}&\ldots&
\vdots\\\vdots&\ldots&
\ddots&\vdots\\0&\ldots&0&\rho^{(0,M)}_{\cal \bar{S}}\end{pmatrix}\,,
\label{blck}
\end{eqnarray}
where  $(\rho^{(m,l)}_{{\cal S}\bar{\cal S}})_{\bm\alpha\bm\alpha'}
= \langle C^{(m,l)\dag}_{\bm\alpha'}C^{(m,l)}_{\bm\alpha}\rangle$, with $C^{(m,l)\dag}_{\bm\alpha}=C^{(m)\dag}_{\bm\alpha\in{\cal S}}C^{(l)\dag}_{\bar{\bm\alpha}\in\bar{\cal S}}$ 
creating $m$ particles in ${\cal S}$ and $l$ in $\bar{\cal S}$. The blocked form \eqref{blck} is equivalent to the condition 
\begin{equation}
[\hat{\rho}^{(M)},\hat N_{\cal S}]=0\end{equation}
on the $m$-body density operator \eqref{rhomo}, implied by a state fulfilling Eq.\ \eqref{Ns}. 
The  operators $C^{(m,l)\dag}_{\bm\alpha}$ satisfy
\begin{equation}\sum_{\bm\alpha}C^{(m,l)\dag}_{\bm\alpha}C^{(m,l)}_{\bm\alpha}={\textstyle\binom{\hat{N}_{\cal S}}{m}\binom{\hat{N}_{\bar{\cal S}}}{l}}\,.\label{sumc}\end{equation} 
Therefore, each block in \eqref{blck} has a definite trace  
 ${\rm Tr}\,\rho^{(l,m)}=\binom{N_{\cal S}}{m} \binom{N_{\bar{{\cal S}}}}{l}$. For $M=1$ and $2$ we recover the blocks of Eqs.\ \eqref{r1} and \eqref{rho2ss}.

The tensor $\Gamma^{(M)}$  will also be decomposed into $M+1$ 
 blocks $\Gamma^{(m,l)}$, each one generating an exact expansion and Schmidt-like decomposition of  $|\Psi\rangle$, since for each $m,l$ Eq.\ \eqref{sumc}   implies $\sum_{\bm{\alpha}}C^{(m,l)\dag}_{\bm\alpha}C^{(m,l)}_{\bm\alpha}|\Psi\rangle=\binom{N_{\cal S}}{m}\binom{N_{\bar{\cal S}}}{l}|\Psi\rangle$ for states satisfying \eqref{Ns}. Thus, we obtain  
\begin{subequations}\label{gexp}
\begin{eqnarray}
\!|\Psi\rangle&\!=&\!\tfrac{1}{\binom{N_{\cal S}}{m}\binom{N_{\bar{S}}}{l}}\!\!\sum_{\bm\alpha, \bm{\beta}}\Gamma^{(m,l)}_{\bm\alpha\bm\beta}C^{(m,l)\dag}_{\bm\alpha} C^{(N_{\cal S}-m,N_{\bar{\cal S}}-l)\dag}_{\bm\beta}|0\rangle, 
\label{expg1}\;\;\;\;\;\;\\
&=&\!\tfrac{1}{\binom{N_{\cal S}}{m}\binom{N_{\bar{S}}}{l}}\!\sum_{\nu}\sigma^{(m,l)}_\nu A^{(m,l)\dag}_\nu B_\nu^{(N_{\cal S}-m,N_{\bar{\cal S}}-l)\dag}|0\rangle
\,,\label{expg2ml}\;\;\;\;\;\;\;\;\;\;\end{eqnarray}
\end{subequations}
with $\Gamma^{(m,l)}_{\bm\alpha\bm\beta}=\langle 0|C_{\bm\beta}^{(N_{\cal S}-m,N_{\bar{\cal S}}-l)}C_{\bm\alpha}^{(m,l)}|\Psi\rangle$, such that  $C_{\bm\alpha}^{(m,l)}|\Psi\rangle=\sum_{\bm{\beta}}\Gamma^{(m,l)}_{\bm\alpha\bm\beta}
C^{(N_{\cal S}-m,N_{\bar{\cal S}}-l)\dag}|0\rangle$ and  $\rho^{(m.l)}=\Gamma^{(m,l)}\Gamma^{(m,l)\dag}$.  In \eqref{expg2ml} $\sigma^{(m,l)}_\nu$ are the singular values of $\Gamma^{(m,l)}$ and $A^{(m,l)\dag}_\nu$, $B_\nu^{(N_{\cal S}-m,N_{\bar{\cal S}}-l)\dag}$ the normal operators obtained from its SVD. These block-DMs and its eigenvalues can be used  to characterize the $M$-body correlations between particles at ${\cal S}$ and $\bar{\cal S}$.   

The $(m,l)$-body density operator corresponding to such block can be likewise obtained as 
\begin{eqnarray} \hat\rho^{(m,l)}&=&\sum_{\bm\beta}C^{(N_{\cal S}-m,N_{\bar{\cal S}}-l)}_{\bm\beta}|\Psi\rangle\langle\Psi|C^{(N_{\cal S}-m,N_{\bar{\cal S}}-l)\dag}_{\bm\beta}\nonumber\\
&=&\sum_{\bm\alpha,\bm\alpha'}\rho^{(m,l)}_{\bm\alpha\bm\alpha'}C^{(m,l)\dag}_{\bm\alpha}|0\rangle\langle 0|C^{(m,l)}_{\bm\alpha'}\,.\end{eqnarray}
Similarly, reduced measurements based on the operators 
\begin{equation}{\cal M}^{(m,l)}_{\bm\alpha}={\textstyle[\binom{{N}_{\cal S}}{m}\binom{{N}_{\bar{\cal S}}}{l}]}^{-1/2}C^{(m,l)}_{\bm\alpha}\,,\end{equation}
which satisfy  $\sum_{\bm\alpha}{\cal M}^{(m,l)\dag}_{\bm\alpha}{\cal M}^{(m,l)}_{\bm\alpha}=\mathbbm{1}_{N_{\cal S}N_{\bar{\cal S}}}$ on the subspace of states with definite particle number $N$ and subsystem particle number $N_{\cal S}$, become feasible, leading to a direct extension of previous Eqs.\ \eqref{pbet}--\eqref{EntL}. The associated entanglement, i.e.\ the mixedness of each of these blocks, will not increase on average under these measurements, and can then be also considered as a resource in this scenario.

\subsection{Connection with entanglement between distinguishable systems}\label{IIID}}
In the special case $M=N_{\cal S}$, the expansion \eqref{gexp} associated to the first block $(m=M,l=0)$  in \eqref{blck}  becomes  
\begin{subequations}
\begin{eqnarray}
|\Psi\rangle&=&\sum_{\bm\alpha\in{\cal S}, \bm{\beta}\in{\cal \bar{S}}}\Gamma^{(N_{\cal S})}_{\bm\alpha\bm\beta} C^{(N_{\cal S})\dag}_{\bm \alpha}C^{(N_{\bar{\cal S}})\dag}_{\bm\beta}|0\rangle\label{55a} \\
&=&\sum_{\nu}\sigma^{(N_{\cal S})}_\nu A^{(N_{\cal S})\dag}_\nu B_\nu^{(N_{\cal S})\dag}|0\rangle
\,,
\label{55b}\end{eqnarray}\label{55}\end{subequations}
with $\Gamma^{(N_{\cal S})}_{\bm\alpha\bm\beta}\equiv \Gamma^{(N_{\cal S},N_{\bar{\cal S}})}_{\bm\alpha\bm\beta}=
\langle 0|C_{\bm\beta}^{(N_{\cal S})} C_{\bm\alpha}^{(N_{\cal S})}|\Psi\rangle$ 
for  $\bm\alpha\in{\cal S}$, $\bm\beta\in{\bar{\cal S}}$. Eq.\ \eqref{55a}
 is just the {\it standard decomposition of a bipartite state of two distinguishable systems} (the ${N}_{\cal S}$ particles at ${\cal S}$ and the ${N}_{\bar{\cal S}}$ at $\bar{\cal S}$) in terms of local states ($C_{\bm\alpha\in{\cal S}}^{(N_{\cal S})\dag}|0\rangle$ and  $C_{\bm\beta\in\bar{\cal S}}^{(N_{\bar{\cal S}})\dag}|0\rangle$) expressed in second quantized form.  
The $N_{\cal S}$ particles at ${\cal S}$ can be distinguished from  the ${N}_{\bar{\cal S}}$  at $\bar{\cal S}$ since they occupy orthogonal subspaces and have then a distinct quantum number.  The diagonal  representation  \eqref{55b} is  the standard Schmidt decomposition,  with Eq.\ \eqref{psiprod} corresponding to the separable case. 

Accordingly,  all terms in the sums \eqref{55a}-\eqref{55b} are now mutually orthogonal. The associated isospectral  $N_{\cal S}$- and $N_{\bar{\cal S}}$-body   densities,  
\begin{equation}\rho^{(N_{\cal S})}_{\cal S}=\Gamma^{(N_{\cal S})}\Gamma^{(N_{\cal S})\dag}\,,\;\;\rho^{(N_{\bar{\cal S}})}_{\bar{\cal S}}=
\Gamma^{(N_{\cal S})t}\Gamma^{(N_{\cal S})*}\,,\label{rhons} \end{equation}
where $\rho_{\cal S}^{(N_{\cal S})}=\rho^{(m,0)}_{\cal S}$ is the first block in \eqref{blck}, represent  the standard local isospectral density matrices of the $N_{\cal S}$ particles at ${\cal S}$ and those at $\bar{\cal S}$, having the same eigenvalues $\lambda_\nu^{(N_{\cal S})}=(\sigma_\nu^{(N_{\cal S})})^2$ and  satisfying ${\rm Tr}\,\rho^{(N_{\cal S})}_{\cal S}=\binom{N_{\cal S}}{N_{\cal S}}=1$, ${\rm Tr}\,\rho^{(N_{\bar{\cal S}})}_{\bar{\cal S}}=\binom{N_{\bar{\cal S}}}{N_{\bar{\cal S}}}=1$, with 
\begin{subequations}
\begin{eqnarray}
\hat\rho^{(N_{\cal S})}_{\cal S}&=&\sum_{\bm \beta\in\bar{\cal S}}C_{\bm\beta}^{(N_{\bar{\cal S}})}|\Psi\rangle\langle\Psi|C_{\bm\beta}^{(N_{\bar{\cal S}})\dag}\,,\\
\hat\rho^{(N_{\bar{\cal S}})}_{\bar{\cal S}}&=&\sum_{\bm \alpha\in{\cal S}}C_{\bm\alpha}^{(N_{\cal S})}|\Psi\rangle\langle\Psi|C_{\bm\alpha}^{(N_{\cal S})\dag} \,,\end{eqnarray}
\label{57m}
\end{subequations}
the corresponding local reduced states. 
Their common entropy  is  the  standard bipartite entanglement entropy of the $N_{\cal S}$ and $N_{\bar{\cal S}}$ particles:
\begin{equation}
E({\cal S},\bar{\cal S})= S(\hat\rho^{(N_{\cal S})})=S(\hat\rho^{(N_{\bar{\cal S}})})\,.\label{ESBS}
\end{equation}

 We finally notice that for a quantum operation transforming an $N$ particle state $|\Psi_0\rangle$ with support on a sp subspace ${\cal S}_0\equiv \bar{\cal S}$,  to a state $|\Psi\rangle$ with $M=N_{\cal S}$ particles in a subspace ${\cal S}$ orthogonal to $\bar{\cal S}$, and $N_{\bar{\cal S}}=N-M$ particles in $\bar{\cal S}$, satisfying then Eq.\ \eqref{Ns}, the result derived in \cite{GDR.21} for fermions also  holds for bosons: The entropy of the original normalized $M$-body DM $\rho_{0n}^{(M)}$ 
 in $|\Psi_0\rangle$ is an upper bound to the average bipartite entanglement \eqref{ESBS} in the final states (see App.\ \ref{B}). 
 \vspace*{0.5cm}

\section{Examples\label{IV}}
We now discuss some examples  illustrating previous considerations in both boson and fermion systems. We focus on $N$-particle paired states, which arise as ground states (GS)  of  systems with attractive pairing interactions. The latter are well-known to be most relevant in several distinct contexts, from the standard BCS theory of superconductivity \cite{bcs.57} and its extension for describing He$^3$ superfluidity \cite{And.61}, 
to the description of pairing effects in nuclear systems and neutron stars   \cite{RS.80,DHJ.03}, including also ultracold quantum gases \cite{ZK.05}.   BCS-like pairing models for bosons have also been considered \cite{Ov.04,DS.01,Richb.68}. 
Such paired states are strongly correlated, requiring, as is well known, at least 
a particle number violating BCS \cite{bcs.57} or Bogoliubov approach  \cite{RS.80}  for an approximate treatment at the mean field level.
We will here consider  exact results  in finite $N$-particle systems, focusing on  the eigenvalues of the one- and, specially, two-body DM  and on the  associated entropies and bipartite expansions, of  some typical paired states, including the GS of a finite pairing model. 
 
\subsection{Maximally paired states in fermionic and bosonic systems}
We start from the uniform  pair creation operator
\begin{equation}
A^\dag=
\frac{1}{\sqrt{n}}\sum_{k=1}^n c^\dag_kc^\dag_{\bar k}\,,\label{Ap1}
\end{equation}
where $k,\bar{k}$ label $n$ orthogonal sp states belonging to orthogonal subspaces ${\cal S}$ and $\bar{\cal S}$ respectively (e.g.\  
$k, \bar{k}$ may label opposite quasimomentum states) and  $c^\dag_k$, $c^\dag_{\bar k}$ can be either  bosonic or fermionic creation operators. 
It creates a  maximally entangled pair state  $|\Psi_1\rangle=A^\dag|0\rangle=\frac{1}{\sqrt{n}}\sum_k c^\dag_kc^\dag_{\bar{k}}|0\rangle$, 
both  in the sense of leading to  a  {\it maximally mixed} one-body DM $\rho^{(1)}=\mathbbm{1}/n$ for $2$ particles in $2n$ levels, i.e.\ maximal one-body entanglement $E^{(1)}$ in Eq.\ \eqref{EntL} for $L=1$, as well as maximum bipartite entanglement  between the two particles (which occupy orthogonal subspaces), i.e.\ 
maximally mixed $\hat\rho^{(N_{\cal S})}$,   $\hat\rho^{(N_{\bar{\cal S}})}$ in \eqref{57m} 
and hence maximal $E({\cal S},\bar{\cal S})$ in  \eqref{ESBS}, for ${N}_{\cal S}=N_{\bar{\cal S}}=1$.    

The operator \eqref{Ap1}  fulfills the commutation relation (in what follows $+$ corresponds to bosons, $-$ to fermions) 
\begin{eqnarray}[A,A^\dag]&=&1\pm \hat{N}/n\,,\label{conm}
\end{eqnarray}
 where  $\hat N=\sum_k c^\dag_k c_k+c^\dag_{\bar k}c_{\bar k}=\hat N_{\cal S}+\hat N_{\cal \bar{\cal S}}$ is the total number operator. 
   Using \eqref{conm} it is straightforward  to show that the ensuing normalized $m$-pair state created by $(A^\dag)^m$ is 
 \begin{subequations}
\begin{eqnarray}
|\Psi_{m}\rangle&:=&\frac{1}{m!}\sqrt{\frac{n^m}{{\cal N}_m}}(A^\dag)^m|0\rangle\label{p1}\\&=&\frac{1}{\sqrt{{\cal N}_{m}}}\sum_{m_1,\ldots,m_n\atop 
\sum_k m_k=m}
|m_1,\ldots,m_n\rangle\,,\label{p2}
\end{eqnarray}
\label{58}
\end{subequations}
$\!\!\!$where $m_k$ is the number of pairs in states $(k,\bar k)$, 
with $m_k=0,1,2,\ldots$ for bosons and $m_k=0,1$  for fermions. In \eqref{p2},   ${\cal N}_m=\binom{n+m-1}{m}$ (bosons) or  ${\cal N}_m=\binom{n}{m}$ (fermions)  is  the number of ways of distributing $m$ indistinguishable pairs in $n$ pair states (with single occupancy in the fermion case and $m\geq 0$ for bosons, $0\leq m\leq n$ for fermions), and    
\begin{eqnarray}
|m_1,\ldots,m_n\rangle=
\prod_{k=1}^n \tfrac{(c^\dag_{k} c^\dag_{\bar k})^{m_k}}{m_k!}|0\rangle=s_{\bm\alpha}
C^{(m)\dag}_{\bm\alpha}C^{(m)\dag}_{\bm{\bar\alpha}}|0\rangle\label{p2m}\,,\;\;\end{eqnarray}
are basic normalized $m$-pair states, with $C^{(m)\dag}_{\bm\alpha}
=\prod_k \frac{(c^\dag_k)^{m_k}}{\sqrt{m_k!}}$, $C^{(m)\dag}_{\bm{\bar\alpha}}= \prod_k \frac{(c^\dag_{\bar k})^{m_k}}{\sqrt{m_k!}}$ for $\bm\alpha=(m_1,\ldots,m_n)$ and $s_{\bm\alpha}=\pm$ a phase factor for the fermionic case.  

 Hence, the states \eqref{58} are just a  {\it uniform superposition} of  these  ${\cal N}_m$  basic 
$m$-pair states, satisfying Eq.\ \eqref{Ns},   
\begin{equation} \hat N_{\cal S}|\Psi_{m}\rangle=\hat{N}_{\bar{\cal S}}|\Psi_{m}\rangle=m|\Psi_{m}\rangle\,.\label{Nsk}
\end{equation}
They arise as exact GS of simple pairing Hamiltonians in the strong coupling limit (see section \ref{C}). 
From \eqref{conm}-\eqref{58} it can be shown that 
\begin{subequations}\label{lad}\begin{eqnarray} A^\dag|\Psi_{m-1}\rangle&=&\sqrt{m(1\pm \tfrac{m-1}{n})}
|\Psi_{m}\rangle\,,\label{lad1}\\
A^\dag A|\Psi_{m}\rangle&=&m(1\pm\tfrac{m-1}{n})|\Psi_{m}\rangle\,,\label{lad2}\end{eqnarray}
\end{subequations}
with $A|\Psi_m\rangle=\sqrt{m(1\pm \frac{m-1}{n})}|\Psi_{m-1}\rangle$, such that 
 $A^\dag A$ counts essentially the number $m$ of pairs for $n\gg m$.

The states \eqref{58} have again {\it maximum one-body}  entanglement (for fixed $N=2m$ particles) for both fermions and bosons:  Eq.\ \eqref{Nsk} implies $\rho^{(1)}$ has the blocked form \eqref{r1}, with $\rho^{(1)}_{\cal S}$, $\rho^{(1)}_{\bar{\cal S}}$ diagonal and {\it maximally mixed}, as all sp states  $k,\bar k$ have the same occupation: 
\begin{equation}\rho^{(1)}_{\cal S}=\rho^{(1)}_{\bar{\cal S}}=\lambda^{(1)}\,\mathbbm{1}\,,\;\;\;
\lambda^{(1)}=m/n\,,\label{la0}\end{equation}
i.e.\  $\langle c^\dag_k c_{k'}\rangle=\langle c^\dag_{\bar k}c_{\bar k'}\rangle=\delta_{kk'}\lambda^{(1)}$, 
verifying ${\rm Tr}\,\rho^{(1)}_{\cal S}={\rm Tr}\,\rho^{(1)}_{\bar{\cal S}}=\frac{1}{2}{\rm Tr}\,\rho^{(1)}=m$ and leading to maximum $E^{(1)}$ in  \eqref{EntL}, i.e.\ $E^{(1)}=\log_2(2n)$ for the von Neumann entropy. 

Similarly, the states \eqref{58}  have also {\it maximum} bipartite  entanglement $E({\cal S},\bar{\cal S})$ in Eq.\ \eqref{ESBS},  between the $N_{\cal S}=m$ particles in ${\cal S}$ and the $m$ ones in  $\bar{\cal S}$, for both fermions and bosons:  
Eqs.\ \eqref{p2}-\eqref{p2m} are already the Schmidt decomposition for such partition,   
$|\Psi_m\rangle=\frac{1}{\sqrt{{\cal N}_m}}
\sum_{\bm\alpha}s_{\bm\alpha}C^{(m)\dag}_{\bm\alpha}C^{(m)\dag}_{\bar{\bm\alpha}}|0\rangle$ with $|s_{\bm\alpha}|=1$, hence leading to  maximally mixed reduced densities $\rho^{(m)}_{\cal S}=\rho^{(m)}_{\bar{\cal S}}={\cal N}_m^{-1}\mathbbm{1}$ and maximum entanglement entropy  
\begin{equation}E({\cal S},\bar{\cal S})=\log_2\,{\cal N}_m.\end{equation}

On the other hand, the  two-body DM $\rho^{(2)}$ determined by the state \eqref{58} is  not maximally mixed, as can be seen from its eigenvalues $\lambda_i^{(2)}$: 
Eq.\ \eqref{Nsk} ensures it will have the blocked structure of Eq. \eqref{rho2ss}, with still  maximally mixed diagonal blocks $\rho^{(2)}_{\cal S}$, $\rho^{(2)}_{\bar{\cal S}}$ (of length $n(n\pm 1)/2)$,
\begin{equation}
\rho^{(2)}_{\cal S}=\rho^{(2)}_{\bar{\cal S}}=\lambda_2^{(2)}\mathbbm{1}\,,\;\;\;\lambda_2^{(2)}=\frac{m(m-1)}{n(n\pm 1)}\,,
\label{la2}
\end{equation}
 since   $\langle c^\dag_{k}c^\dag_{k'}c_{k'''}c_{k''}\rangle=
\delta_{kk''}\delta_{k'k'''}\lambda_2^{(2)}$   for $k<k'$, $k''<k'''$, 
and additionally  $\langle \frac{c^{\dag\,2}_{k}}{\sqrt{2}}
\frac{c_{k'}^2}{\sqrt{2}}\rangle=\delta_{kk'}\lambda_2^{(2)}$ for bosons (see App.\  \ref{A}), with identical expressions in $\bar{\cal S}$,  verifying  ${\rm Tr}\,\rho^{(2)}_{\cal S}={\rm Tr}\,\rho^{(2)}_{\bar{\cal S}}=\binom{m}{2}$ for bosons and fermions. 

However, the  remaining block $\rho^{(2)}_{{\cal S}\bar{\cal S}}$ in \eqref{rho2ss}, of length $n^2$ for bosons and fermions, becomes itself blocked in two submatrices  (see  Eq.\ (A14) in App.\ \ref{A}), 
\begin{subequations}
\begin{eqnarray}\langle c^\dag_{k}c^\dag_{\bar{k}'}c_{\bar{k}'''}c_{k''}\rangle\!&=&\!(1\!-\!\delta_{kk'})\delta_{kk''} \delta_{k'k'''}\lambda_2^{(2)} \label{61a}\\&&+\delta_{kk'}\delta_{k''k'''}[\tfrac{m(n\pm m)}{n(n\pm 1)}+\delta_{kk''}\lambda^{(2)}_2],\;\;\;\;\;\;\;\;\label{61b}\end{eqnarray}
\label{70}
\end{subequations}
$\!$where the first block  \eqref{61a}  is diagonal and similar to \eqref{la2}, while the second block \eqref{61b},  of length $n$, is nondiagonal and exposes  the two-body pairing correlations between particles in ${\cal S}$ and $\bar{\cal S}$.  
It has two distinct eigenvalues: one given again by $\lambda^{(2)}_2$,  Eq.\ \eqref{la2}, 
  $n-1$ degenerate in this sub-block, while the remaining one (see App.\ \ref{A}),  
 \begin{eqnarray}
\lambda_1^{(2)}&=&m\left(1\pm\frac{m-1}{n}\right)\,,\label{la1}
\end{eqnarray}
 is the {\it single  dominant nondegenerate} eigenvalue of $\rho^{(2)}$, satisfying  $\lambda_1^{(2)}\geq m$ for bosons  
and $\lambda_1^{(2)}\geq 1$ for fermions 
(with $\lambda_1^{(2)}>m$ for bosons if $m>1$ and $\lambda_1^{(2)}>1$ for fermions if $1<m<n$). It  corresponds to the flat normal operator    $A_1^{(2)\dag}=A^\dag$ of $\rho^{(2)}$, 
as  $A^\dag A|\Psi_m\rangle=\lambda^{(2)}_1|\Psi_m\rangle$ (Eq.\   \eqref{lad2}) and hence $\langle A^\dag A\rangle=\lambda^{(2)}_1$. 
 
Thus, for $n\gg m$, $\lambda_1^{(2)}\approx m$ is essentially the number $m$ of pairs  while $\lambda_2^{(2)}\approx (m/n)^2$ becomes small, in agreement with the approximate bosonic interpretation of $A^\dag$ for $n\gg N$ (Eq.\ \eqref{conm}), in which case the state \eqref{p1} can be seen as an $m$-boson condensate. Nonetheless, their exact values are required for fulfilling Eqs.\ \eqref{tr2}--\eqref{tr2ssb}: ${\rm Tr}\,\rho^{(2)}=\lambda^{(2)}_1+(n(2n\pm 1)\!-\!1)\lambda^{(2)}_2=\binom{2m}{2}$,  
  ${\rm Tr}\,\rho^{(2)}_{\cal S\bar{\cal S}}=\lambda^{(2)}_1\!+\!(n^2\!-\!1)\lambda^{(2)}_2=m^2$, and are important when $m\sim n$ or $m>n$ (bosonic case). 
In the fermionic case Eq.\ \eqref{la1} is also  
 the {\it largest} value the maximum  eigenvalue of $\rho^{(2)}$ can reach among any state of $N=2m$ particles in a $2n$-dimensional sp space  \cite{Yang.62,Samb.20}. 

We can now verify expansions \eqref{rexp} and \eqref{e1}--\eqref{e3}. Blocks  $\rho^{(1)}_{\cal S}$ and $\rho^{(1)}_{\bar{\cal S}}$  generate similar  $(1,N-1)$ uniform expansions  \eqref{40b}  of $|\Psi_{m}\rangle$: From \eqref{p1}, 
$c_k (A^\dag)^m|0\rangle=\frac{m}{\sqrt{n}}c^\dag_{\bar k}(A^\dag)^{(m-1)}|0\rangle$, such that  \eqref{17b} is verified for $M=1$,  $A^{(1)}_\nu=c_k$:    $c_k|\Psi_m\rangle=
\sqrt{\lambda^{(1)}} B^{(2m-1)\dag}_k|0\rangle$, with $B^{(2m-1)\dag}_k|0\rangle= 
\frac{c^\dag_{\bar k}}{\sqrt{1\pm \frac{m-1}{n}}}|\Psi_{m-1}\rangle$   
 the normalized state of remaining  particles after one in sp state $k$ is annihilated.  Eq.\ \eqref{40b} is then  fulfilled:  $\frac{1}{m}\sqrt{\frac{m}{n}}\sum_k c^\dag_k B^{(2m-1)\dag}_k|0\rangle=\frac{A^\dag |\Psi_{m-1}\rangle}{\sqrt{m(1\pm \frac{m-1}{n})}}=|\Psi_m\rangle$, according to \eqref{lad1}. 

 Likewise, blocks $\rho^{(2)}_{\cal S}$,  $\rho^{(2)}_{\bar{\cal S}}$ generate  similar  uniform  expansions \eqref{e1} of $|\Psi_{m}\rangle$: for $k\neq k'$,   $c_{k'}c_{k}(A^\dag)^m|0\rangle$ $=\frac{m(m-1)}{n}c^\dag_{\bar k'}c^\dag_{\bar k}(A^{\dag})^{(m-2)}|0\rangle$ and hence $c_{k'}c_k|\Psi_m\rangle=  \sqrt{\lambda^{(2)}_2}B_{k'k}^{(2m-2)\dag}|0\rangle$, 
  where $B_{k'k}^{(2m-2)\dag}|0\rangle=\frac{c^\dag_{\bar{k}'}c^\dag_{\bar k}|\Psi_{2m-2}\rangle}{\langle c_{\bar k}c_{\bar k'}c^\dag_{\bar{k}'}c^\dag_{\bar k}\rangle^{1/2}}$ is the normalized state of remaining  particles after annihilating two particles in  states $k,k'$.  Eq.\ \eqref{e1} is  then verified: $\frac{\sqrt{\lambda^{(2)}_2}}{m(m-1)}\sum_{k,k'}c^\dag_{k}c^\dag_{k'} B_{k'k}^{(2m-2)\dag}|0\rangle=|\Psi_m\rangle$. 
  
  On the other hand the expansion \eqref{e3} based on $\rho^{(2)}_{{\cal S}\bar{\cal S}}$ has here a {\it single dominant term:}  Eq.\ \eqref{lad1} is just  \eqref{17b} for the main  eigenvalue $\lambda^{(2)}_1$ and eigenvector $A^{(2)}_1=A$:  $A|\Psi_m\rangle=\sqrt{\lambda^{(2)}_1}|\Psi_{m-1}\rangle$,  with  $B^{(N-2)\dag}_1\propto (A^\dag)^{m-1}$. Moreover,  the first term alone in \eqref{e3b} 
  is  already proportional to the {\it exact} state, as $A^{\dag}A|\Psi_m\rangle\propto |\Psi_m\rangle$ according to \eqref{lad2}. The sum of all remaining terms in \eqref{e3b} is in this case   proportional to this first term. 

\subsection{General paired states}
Let us now consider  the general $m$-pair state 
\begin{subequations}
\begin{eqnarray} |\Psi_m\rangle&=&\sum_{\bm\alpha}\Gamma_{\bm\alpha}C^{(m)\dag}_{\bm\alpha}C^{(m)\dag}_{\bm{\bar\alpha}}|0\rangle\\
&=&\sum_{m_1,\ldots,m_n\atop \sum_k m_k=m}\Gamma_{m_1\ldots m_n}|m_1,\ldots,m_n\rangle\label{70b}
\end{eqnarray}\label{GS}
\end{subequations}
$\!\!$where $|m_1,\ldots,m_n\rangle$ are the previous states \eqref{p2m} and $\Gamma_{\bm\alpha}=\Gamma_{m_1\ldots m_n}$ arbitrary coefficients satisfying $\sum_{\bm\alpha}|\Gamma_{\bm\alpha}|^2=1$, with $m_k=0,1,2,\ldots$ ($0,1$) for bosons (fermions). Like \eqref{58}, these states contain all $N=2m$ particles in $m$ pairs $(k,\bar{k})$ and arise as GS of pairing Hamiltonians at finite couplings strengths (see sec.\ \ref{C}). They  satisfy Eqs.\ \eqref{Ns}-\eqref{Nsk}, then leading to the same  blocked structure 
\eqref{r1}--\eqref{rho2ss} of $\rho^{(1)}$ and $\rho^{(2)}$ for fermions and bosons, with $\rho^{(1)}$ and $\rho^{(2)}_{\cal S}$, $\rho^{(2)}_{\bar{\cal S}}$ again diagonal in the standard basis: 
\begin{subequations}
\begin{eqnarray}
\langle c^\dag_k c_{k'}\rangle=\langle c^\dag_{\bar k}c_{\bar k'}\rangle&=&\delta_{kk'}\lambda^{(1)}_k\,,\label{71a}\\
\langle c^\dag_{k} c^{\dag}_{k'}c_{k'''}c_{k''}\rangle&=&\delta_{kk''}\delta_{k'k'''}\lambda^{(2)}_{kk'}\,,\label{71b}
\end{eqnarray}
\label{71}
\end{subequations}
$\!$for $k<k'$,  $k''<k'''$, and similarly  for $k\rightarrow \bar k$, with $\lambda^{(2)}_{\bar k\bar k'}=\lambda^{(2)}_{kk'}$ and $\lambda^{(2)}_{kk}=
\frac{1}{2}\langle c^{\dag 2}_kc_k^2\rangle=\lambda^{(2)}_{\bar k  \bar k}$
for bosons. And for $\rho^{(2)}_{\cal S \bar{\cal S}}$,  Eq.\ \eqref{70} is replaced by 
\begin{subequations}
\begin{eqnarray}\langle c^\dag_{k} c^\dag_{{\bar k}'}c_{{\bar k}'''}c_{k''}\rangle&=&(1-\delta_{kk'})\delta_{k k''}
\delta_{k'k'''}\lambda^{(2)}_{kk'}
\label{72a}\\&&+\delta_{kk'}\delta_{k''k'''}\rho^{(2)}_{c\,kk''}\label{72b}\end{eqnarray}\label{72}\end{subequations}
 such that $\rho^{(2)}_{{\cal S}\bar{\cal S}}=\left(^{\rho^{(2)}_d\;0}_{\;0\;\;\rho^{(2)}_c}\right)$, with $\rho^{(2)}_d$ the diagonal sub-block \eqref{72a} having the same elements \eqref{71b},  and $\rho^{(2)}_c$ the nondiagonal $n\times n$ ``collective'' sub-block \eqref{72b} containing the 
two-body pairing contractions  $\langle c^\dag_kc^\dag_{\bar k}c_{\bar k'}c_{k'}\rangle$. 

In the {\it fermionic} case, this  sub-block yields itself to an exact $(2,N\!-\!2)$ reduced expansion of $|\Psi_m\rangle$ containing at most $n$ terms, since 
$\hat{N}_p\equiv\sum_k c^\dag_k c^\dag_{\bar k}c_{\bar k} c_k$ just counts for fermions the number of pairs, satisfying $\hat N_p|\Psi_m\rangle= m|\Psi_m\rangle$. Thus,   $|\Psi_m\rangle=\frac{1}{m}\hat{N}_p|\Psi_m\rangle$ can be expanded as 
\begin{subequations}\begin{eqnarray}
|\Psi_m\rangle&=&\!\tfrac{1}{m}\!\sum_{k,\bm\beta}
\Gamma^{(2)}_{k\bm\beta} c^\dag_k c^\dag_{\bar k}B^{(N-2)}_{\bm\beta}|0\rangle\\
&=&\!\tfrac{1}{m}\!\sum_{\tilde\nu}\sigma^{(2)}_{\tilde\nu} A^{(2)\dag}_{\tilde\nu} B^{(N-2)\dag}_{\tilde\nu}|0\rangle,
\;\;\;\;\;\;\label{73b}
\end{eqnarray}
\label{73}
\end{subequations}
$\!\!$where we have written $c_{\bar k}c_k|\Psi_m\rangle=\sum_{\bm\beta}\Gamma^{(2)}_{k\bm\beta} B^{(N-2)\dag}_{\bm\beta}|0\rangle$ with $\Gamma^{(2)}_{k\beta}=\langle 0|B^{(N-2)}_{\bm\beta}c_{\bar k}c_k|\Psi_m\rangle$ the elements of the ``collective'' sub-block $\Gamma_c^{(2)}$ of the full $\Gamma^{(2)}_{{\cal S}\bar{\cal S}}$  in \eqref{e3}, such that $\Gamma^{(2)}_c\Gamma^{(2)\dag}_c=\rho^{(2)}_c$ in \eqref{72}. 
In \eqref{73b} $\sigma^{(2)}_\nu$ are the singular values of this sub-block, with  $\lambda^{(2)}_\nu=(\sigma_\nu^{(2)})^2$ the eigenvalues of $\rho^{(2)}_c$ and   $A^{(2)\dag}_\nu,B^{(N-2)\dag}_\nu$ the associated normal operators determined by its SVD. In the presence of a dominant eigenvalue, a good approximation to $|\Psi_m\rangle$ can be obtained with just a few terms in \eqref{73} (see \ref{C}). 

In particular, for a  pair creation operator of the form 
\begin{equation}{A}^{\dag}=
\sum_{k=1}^m \sigma_k c^\dag_k c^\dag_{\bar k}\label{Ad}\,,\end{equation}
where $\sum_k|\sigma_k|^2=1$ and we can set $\sigma_k$ real  $\geq 0$ by adjusting the phases of the $c^\dag_k$, an example of \eqref{GS} is 
\begin{equation}\begin{split}|\Psi_m\rangle&= \frac{1}{m!\sqrt{{\cal N}'_m}}({A}^\dag)^m|0\rangle\\
&=
\frac{1}{\sqrt{{\cal N}'_m}}\sum_{m_1,\ldots,m_n\atop \sum_k m_k=m}\!\!\!\!\sigma_1^{m_1}\ldots\sigma_n^{m_n}|m_1,\ldots,m_n\rangle\,,\label{psimp}
\end{split}
\end{equation}
where 
 ${\cal N}'_m
=\sum_{m_1,\ldots,m_n}\sigma_1^{2m_1}\ldots,
\sigma_n^{2m_n}$ (with the same previous restrictions on the $m_k$ for bosons or fermions). These states are just   particle number projected BCS-like states: $|\Psi_m\rangle\propto P_m|{\rm BCS}\rangle$, 
where $|{\rm BCS}\rangle\propto \exp[A^\dag]|0\rangle=\prod_k\exp(\sigma_k c^\dag_k c^\dag_{\bar k})|0\rangle$ \cite{RS.80,DGR.18}  ($=\prod_k(1+\sigma_k c^\dag_k c^\dag_{\bar k})|0\rangle$ for fermions) and $P_{m}$ is the projector onto $m$-pair ($2m$-particle) states. 

We now prove that in {\it all} states \eqref{psimp} (but not \eqref{GS}) the largest  eigenvalue $\lambda^{(2)}_1$ of $\rho^{(2)}_{{\cal S}\bar{\cal S}}$ (stemming from $\rho^{(2)}_c$ in 
\eqref{72})  satisfies 
\begin{equation} \lambda^{(2)}_1\geq \left\{\begin{array}{rl}m & ({\rm bosons})\\
 1&({\rm fermions}) \end{array}\right.\label{bound1}\,,
\end{equation}
for {arbitrary} $\{\sigma_k\}$: 
A straightforward evaluation of the average  in the state \eqref{psimp} yields (see (A.15)-(A.16)), 
\begin{subequations}
\label{Fg}\begin{eqnarray}
\langle {A}^\dag A\rangle&=&
m\pm (m\!-\!1){\textstyle\sum_k}\sigma_k^2
\langle c^{\dag}_kc_k\rangle\label{Fga}\\
&=&1+(m\!-\!1)(1\pm {\textstyle\sum_k}\sigma_k^2\langle c^\dag_k c_k\rangle)\,,\;\;\;\;\;\;\;
\label{Fgb}
\end{eqnarray}
\end{subequations}
 for bosons ($+$) or fermions ($-$),  such that \eqref{Fga} implies $\langle A^{\dag}A\rangle\geq m$ for bosons and \eqref{Fgb}  $\langle A^{\dag}A\rangle\geq 1$ for fermions (where  $\sum_k\sigma_k^2\langle c^\dag_k c_k\rangle\leq \sum_k\sigma_k^2=1$). This proves Eq.\ \eqref{bound1} since the largest eigenvalue $\lambda^{(2)}_1$ of  $\rho^{(2)}_{{\cal S}\bar{\cal S}}$ should satisfy $\lambda^{(2)}_1\geq \langle {A}^\dag A\rangle$. 
 
 And to see it does not hold for all states \eqref{GS}, just take a superposition of states with orthogonal sp support, e.g.\ $\frac{1}{\sqrt{2}}(\prod_{k=1}^m c^\dag_kc^\dag_{\bar k}+\prod_{k=m+1}^{2m}c^\dag_kc^\dag_{\bar k})|0\rangle$, for which the nonzero eigenvalues of $\rho^{(2)}$ are just all $1/2$ for both bosons and fermions $\forall\,m\geq 2$. \qed 
   
   Eq.\ \eqref{Fg}  also shows that ${A}^\dag A$ has itself a largest eigenvalue $\lambda^N_{\rm max}\geq m\; (1)$ for bosons (fermions) amongst $N=2m$-particle states, since  again $\lambda^N_{\rm max}\geq \langle {A}^\dag A\rangle$ for the average taken in any $N$ particle state.  
Notice, however, that for general $\sigma_k$  the state \eqref{psimp} is no longer an exact eigenstate of ${A}^\dag A$, nor is ${A}^\dag$ a normal mode of the associated $\hat\rho^{(2)}$.  

In the uniform case $\sigma_k=1/\sqrt{n}\,\forall\,k$, $\langle c^\dag_k c_k\rangle=m/n\,\forall\,k$ and we recover from \eqref{Fg} the result \eqref{la1}. Moreover, while (as previously mentioned) in fermion systems this is the {\it maximum}  value $\lambda^{(2)}_1$  can reach among  all $2m$-particle states in a $2n$-dimensional sp space, in boson systems Eq.\ \eqref{la1} represents actually  the {\it minimum}  value reached by the maximum eigenvalue $\lambda_1^{(2)}$  among the states \eqref{psimp}: Since for nonuniform $\sigma_k>0$, $\langle c^\dag_k c_k\rangle>\langle c^\dag_{k'} c_{k'}\rangle$ if 
$\sigma_k>\sigma_{k'}$ while $\sum_k \langle c^\dag_k c_k\rangle =m$ and $\sum_k\sigma_k^2=1$ are fixed, we obtain,  writing $\langle c^\dag_kc_k\rangle=m/n+\delta_k$, with $\sum_k \delta_k=0$,  the bound 
$\sum_k \sigma_k^2\langle c^\dag_k c_k\rangle\geq m/n$, such that \eqref{Fga} and previous fermionic result imply, for all states \eqref{psimp},   
\begin{equation} 
\begin{array}{rl}\lambda^{(2)}_1\geq m(1+\frac{m-1}{n})\; & ({\rm bosons})\\\lambda^{(2)}_1 
 \leq m(1-\frac{m-1}{n})\;&({\rm fermions}) \end{array}\label{bound2}\,,
\end{equation}
which for bosons is stronger than  \eqref{bound1}.  Maximum $\lambda^{(2)}_1$ in the bosonic case amongst the states \eqref{psimp} is obtained when all pairs are in a single state $k$ ($\lambda^{(2)}_1=m^2$). 

Finally, we recall that in the fermion case {\it any} pair creation operator  $A^\dag=\frac{1}{2}\sum_{i,j}\Gamma_{ij}c^\dag_i c^\dag_j$ (with $\Gamma_{ji}=-\Gamma_{ij}$) can be written in the previous normal form \eqref{Ad} 
 \cite {ES.02} (directly related  to the normal form  \eqref{ScDcx} for $N=2$, $M=1$ of the state $A^\dag|0\rangle$), with 
 $\sigma_k$ the singular values of  $\Gamma$  and the $c^\dag_k, c^\dag_{\bar k}$, unitarily related to the $c^\dag_i$.  

\subsection{Finite pairing system\label{C}}
We finally consider the exact ground state (GS) of a finite discrete pairing model. Such model describes finite superconducting systems in the fermionic case (see e.g.\ \cite{RS.80,vondelft,RCR.98}), while its bosonic version has also been considered \cite{Ov.04,DS.01,Richb.68}. Studies of entanglement in such systems  have mainly focused on mode-type entanglement in the  approximate BCS GS   \cite{PV.14,DCL.05,CL.08,OK.05} or on the fermionic one-body entanglement and concurrence  \cite{DGR.18}. Here we will  concentrate on  the two-body entanglement determined by $\rho^{(2)}$ and the associated state expansions, in both the fermionic and bosonic version of the model. 

As in previous examples, we will work within an effective  single particle subspace of dimension $2n$, spanned by $n$ states $k$ and $n$ states  $\bar k$, 
with sp levels of energies $\varepsilon_k=\varepsilon_{\bar k}$. The Hamiltonian is 
\begin{equation}
    H = \sum_k \varepsilon_k (c_k^\dagger c_k + c_{\bar{k}}^\dagger c_{\bar{k}}) - \sum_{k,k'} G_{k k'}c_{k'}^\dagger c_{\bar{k'}}^\dagger c_{\bar{k}} c_k\,,
    \label{Hp}
\end{equation}
where the second term is the pairing interaction. 
 It  conserves the number of particles in states $k$ (subspace ${\cal S}$) and   $\bar{k}$ (subspace $\bar{\cal S}$), satisfying 
\begin{equation}
[H,N_{\cal S}]=[H,N_{\bar{\cal S}}]=0\,. 
\end{equation}
Hence its exact eigenstates, and in particular its GS,  will satisfy Eqs.\ \eqref{Ns}-\eqref{Nsk}. For  even $N=2m$ and $G_{kk'}>0$  $\forall k,k'$, the exact GS will be of the  form \eqref{GS} for both bosons and fermions,  since in order to minimize its energy it will have all $N$ particles in $m$ pairs $(k,\bar k)$, without  broken pairs.  

In what follows we consider a constant sp spacing $\varepsilon_{k+1}-\varepsilon_k=\varepsilon$ $\forall k$ and uniform coupling strength $G_{kk'} = G\geq 0$ $\forall\,k,k'$, such that the interaction in \eqref{Hp} becomes $nG A^\dag A$ with  $A^\dag$  the uniform pair creation operator  \eqref{Ap1}. 

Thus, for $g\equiv G/\varepsilon\rightarrow\infty$, the GS of $H$ will approach that of $-nGA^\dag A$, which is the maximally paired 
state $|\Psi_m\rangle\propto (A^\dag)^m|0\rangle$, Eq.\  \eqref{58}, as it  maximizes $\langle A^\dag A\rangle$ for any fixed $N=2m$. 
For a uniform spectrum centered at $0$, $\varepsilon_k=\varepsilon(k-\frac{n+1}{2})$, $k=1,\ldots,n$,  the energy 
$E_m=\langle\Psi_m|H|\Psi_m\rangle$ of such state is  (Eqs.\ \eqref{lad2}-\eqref{la1}) 
\begin{equation} E_{m}=-nG\lambda^{(2)}_1=-mG[n\pm (m-1)]\,,\label{res21}\end{equation}
where $+$ ($-$) is for bosons (fermions). 

On the other hand, for $g\rightarrow 0^+$ the GS will approach  $|\Psi_m^0\rangle=|m,0,\ldots,0\rangle$ for bosons, $|\Psi_m^0\rangle=|1_1\ldots 1_{m},0,\ldots 0\rangle$ for fermions (in terms of the paired states \eqref{p2m}), with  $E^0_m=\langle\Psi^0_m|H|\Psi^0_m\rangle= -m[\varepsilon(n-a)+bG]$ and $a=1\;(m)$, $b=m\;(1)$ for bosons (fermions). Therefore,  $E_m-E_m^0=m(\varepsilon-G)(n-a)<0$ already for $G>\varepsilon$. 
The exact GS  for finite $n,m$  
 will then evolve  continuously  from $|\Psi^0_m\rangle$ to the state \eqref{58}  as $g$ increases from $0$ 
 to $\infty$, through states of the form \eqref{GS}.   
 
\subsubsection{Fermionic system}
We have  analyzed in \cite{DGR.18} the one-body entanglement determined by $\rho^{(1)}$ in the fermionic version of this system, together with the fermionic concurrence of the reduced state of 4 modes and other related aspects.  
 We will here focus on the two-body DM and the associated entanglement entropy and exact expansions of the GS.  

\begin{figure}
     \includegraphics[scale=.35
     %,trim={6.cm 2.25cm 6cm 2.5cm},clip
     ]{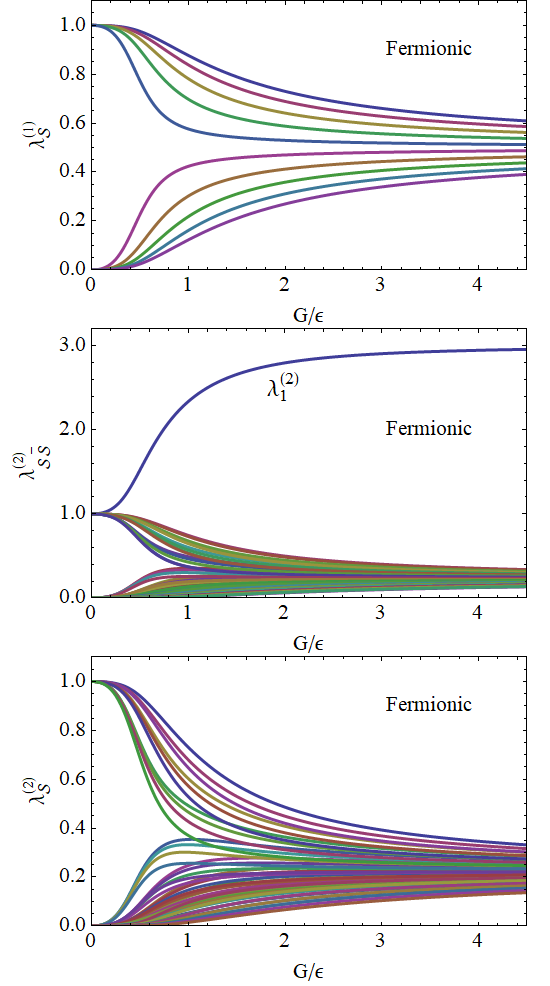}     \vspace*{-0.3cm}
     
     \caption{Eigenvalues of the one-body (top) and two-body (center and bottom) density matrices $\rho^{(1)}$ and $\rho^{(2)}$ as a function of the scaled coupling strength $g=G/\varepsilon$ (dimensionless) in the GS of the  Hamiltonian \eqref{Hp} for a finite half-filled fermionic case ($n=10$). The central panel depicts those of the central block $\rho^{(2)}_{{\cal S}\bar{\cal S}}$ in \eqref{rho2ss}, containing the dominant eigenvalue  $\lambda^{(2)}_1\geq 1$,  while the bottom panel those  of $\rho^{(2)}_{\cal S}=\rho^{(2)}_{\bar{\cal S}}$.}
     \vspace*{-0.35cm}
\label{fig1}\end{figure}

We first depict in Fig.\ \ref{fig1} the exact eigenvalues (i.e.\ the entanglement spectrum) of the one- and two-body DMs in the fermionic case as a function of $g=G/\varepsilon$. We have considered a half-filled system $N=2m=n$, with $n=10$. At $g=0$ the GS $|\Psi_0^m\rangle$ has just the bottom half levels occupied, such that all eigenvalues of $\rho^{(1)}$ and $\rho^{(2)}$ start from $1$ or $0$ at $g=0$.  Both the eigenvalues $\lambda^{(1)}_k$ of $\rho^{(1)}_{\cal S}$  (top panel)   and  $\lambda^{(2)}_{kk'}$ of  $\rho^{(2)}_{\cal S}$ (bottom panel),  Eq.\ \eqref{71},    identical to those of  $\rho^{(1)}_{\bar{\cal S}}$, $\rho^{(2)}_{\bar{\cal S}}$, become ``more mixed'' and $<1$ as the coupling $g$ increases, reflecting the departure of the GS from a SD as all levels above the Fermi level start to be occupied. They  exhibit maximum variation  
around the transition  region $g\approx 1$ 
and approach the maximally mixed limit (for such $N$) for $g\rightarrow\infty$, where they all coalesce with the values  \eqref{la0}-\eqref{la2}, i.e.,   $\lambda_k^{(1)}\rightarrow 1/2$, 
 $\lambda^{(2)}_{kk'}\rightarrow \frac{1}{4}\frac{1-2/n}{1-1/n}\approx\frac{1}{4}(1-\frac{1}{n})$ in the half-filled case.   These results imply a monotonously increasing one- and  two-body entanglement within ${\cal S}$ for increasing pairing strength, saturating for $g\rightarrow\infty$. 

\begin{figure}
     \includegraphics[scale=.35
     %,trim={6.cm 2.25cm 6cm 2.5cm},clip
     ]{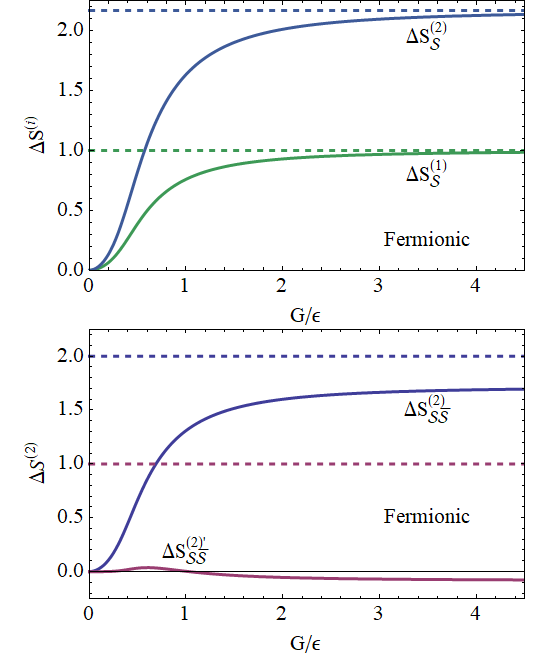}     \vspace*{-0.3cm}
     
     \caption{The entropy increment  \eqref{DS} 
     of the normalized one- and two-body  DMs  
     as a function of the scaled strength $g=G/\varepsilon$ in the fermionic case of Fig.\ \ref{fig1}. The top panel depicts results for that of $\rho^{(1)}_{\cal S}$  ($\Delta S^{(1)}_{\cal S}$) and of the two-body block $\rho^{(2)}_{\cal S}$ ($\Delta S^{(2)}_{\cal S})$ in \eqref{rho2ss}, whereas the bottom panel those of the central block $\rho^{(2)}_{{\cal S} \bar {\cal S}}$ and its collective  sub-block $\rho^{(2)}_{c}$ ($\Delta^{(2)'}_{{\cal S}\bar{\cal S}})$, Eq.\ \eqref{72b}.   The dashed lines indicate their maximum values (reached for a maximally mixed DM).}
     \vspace*{-0.25cm}
     \label{fig2}\end{figure}

 In contrast, the two-body DM  block $\rho^{(2)}_{\cal S\bar{\cal S}}$ (central panel) exhibits instead a single {\it dominant eigenvalue} $\lambda^{(2)}_1>1$ $\forall g>1$ which departs from the rest (that behave as those of $\rho^{(2)}_{\cal S}$) and increases for increasing  $g$,  approaching the limit \eqref{la1} ($=\frac{n}{4}+\frac{1}{2}$ in the half-filled case) for $g\rightarrow \infty$.  It is characteristic of pairing correlations and stems from the collective sub-block $\rho^{(2)}_c$, Eq.\ \eqref{72b}, reflecting for large $g$ the ``multiple occupation'' of the collective pair state created by $A^\dag$.  It prevents this block from becoming more mixed as $g$ increases, implying a two-body ${\cal S}\bar{\cal S}$ entanglement below maximum for $g\rightarrow\infty$. 
 
 In Fig.\ \ref{fig2} we depict the associated one- and two-body entropy increments ($\alpha={\cal S}$ or ${\cal S}\bar{\cal S}$, $i=1,2$)
 \begin{equation}
 \Delta S^{(i)}_\alpha=S[\rho^{(i)}_{n\alpha}(g)]-S[\rho^{(i)}_{n\alpha}(0)]\,,\label{DS}\end{equation}
 which quantify the  entanglement generated by the coupling, 
 where $S(\rho)=-{\rm Tr}\,\rho\log_2\rho$ and $\rho_{n\cal \alpha}^{(i)}(g)$ denotes the normalized $i$-body DM of block $\alpha$ at coupling $g$. Accordingly, for $\rho^{(1)}_{\cal S}$ and  $\rho^{(2)}_{\cal S}$ (upper panel) this  difference increases monotonously from $0$ as  $g$ increases, reaching its saturation value for $g\rightarrow\infty$, where  $\Delta^{(1)}_{\cal S}\rightarrow\log_2\frac{2n}{n}=1$,  $\Delta^{(2)}_{\cal S}\rightarrow\log_2[\binom{n}{2}/\binom{n/2}{2}]\approx 2+\frac{1}{n\ln 2}$ in the half-filled case. 

In contrast, for $\rho^{(2)}_{\cal S \bar{\cal S}}$ (bottom), though increasing with $g$, $\Delta^{(2)}_{\cal S\bar \cal S}$ stays below the saturation value $\log_2 \frac{n^2}{(n/2)^2}=2$  due to the dominant eigenvalue $\lambda_1^{(2)}$, reaching for $g\rightarrow\infty$ the lower limit  
$-p\log_2 p-(1-p)\log_2\frac{1-p}{n^2-1}-2\log_2\frac{n}{2}\approx 2- \frac{\log_2 (n/e)}{n}$, where $p=\frac{\lambda_1^{(2)}}{(n/2)^2}$  
with $\lambda_1^{(2)}$ given by \eqref{la1}.

We also depict in the lower panel the entropy increment $\Delta^{(2)'}_{\cal S\bar\cal S}$ of the collective  $n\times n$ sub-block $\rho^{(2)}_{cn}$ of  $\rho^{(2)}_{{\cal S} \bar {\cal S}}$,  containing just the contractions $\langle c^\dag_k c^\dag_{\bar k}c_{\bar k'}c_{k'}\rangle$ (Eq.\ \eqref{72b}) and hence the dominant eigenvalue, which best reflects its effect. This increment  actually becomes {\it negative} for large $g$, approaching $\approx -\frac{1}{2}\log_2 \frac{n}{16}+O(
\tfrac{\log_2 n}{n})$ for large $n$, well below its saturation value $\log_2 \frac{n}{n/2}=1$. Thus, in this limit the entropy of this sub-block becomes {\it lower} than  in the noninteracting case, reflecting the ``separable-like'' $(2,N\!-\!2)$ form of the limit state \eqref{p1}. 
\vspace*{-0.5cm}  
      
      \begin{figure}[t]
     \includegraphics[scale=.35
     %,trim={6.cm 2.25cm 6cm 2.5cm},clip
     ]{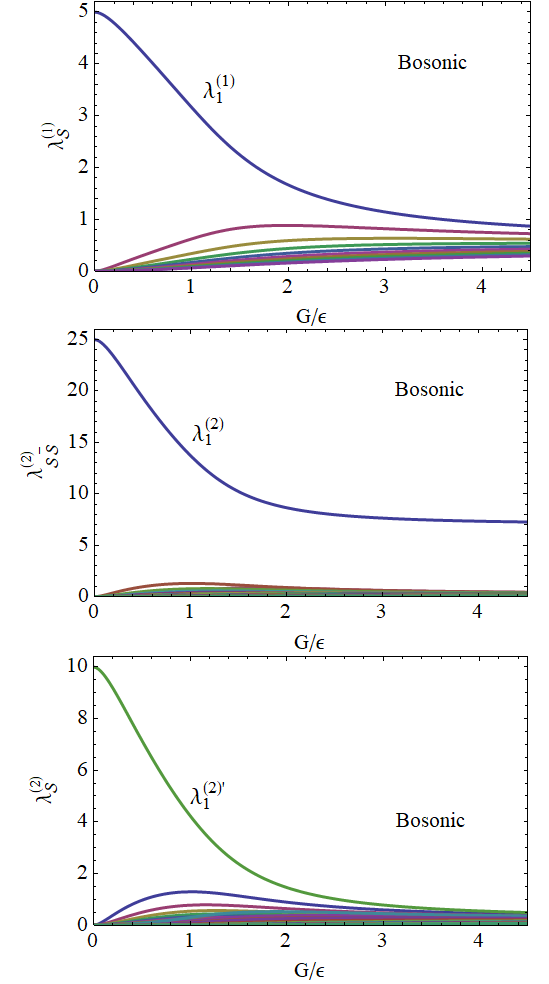}     \vspace*{-0.25cm}
     \caption{Eigenvalues of the one-body (top) and two-body (center and bottom) density matrices $\rho^{(1)}_{\cal S}$ and $\rho^{(2)}_{{\cal S}\bar{\cal S}}$, $\rho^{(2)}_{\cal S}$ as a function of the scaled strength $G/\varepsilon$ in the GS of  Hamiltonian \eqref{Hp} for the  $N=n$ bosonic case. Details are similar to those of Fig.\ \ref{fig1}). The dominant eigenvalue of each block is indicated.}
     \vspace*{-0.35cm}
 \label{fig3}\end{figure}

   \subsubsection{Bosonic system}    
\begin{figure}
     \includegraphics[scale=.35
     %,trim={6.cm 2.25cm 6cm 2.5cm},clip
     ]{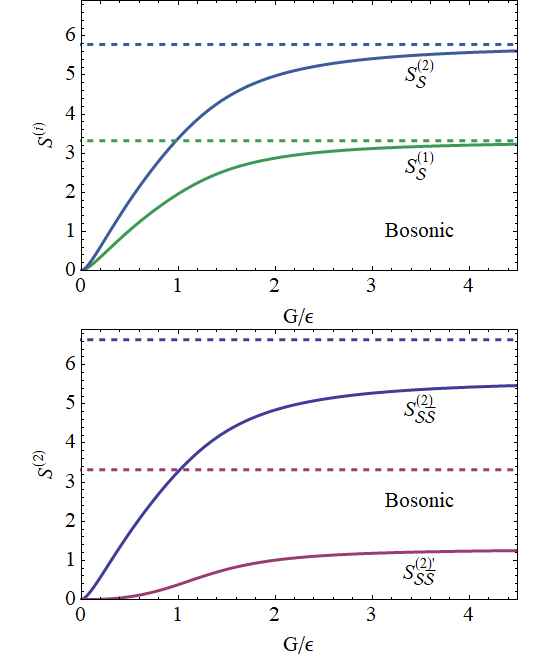}  \vspace*{-0.25cm}
     
     \caption{The entropies $S^{(i)}=S(\rho_{{\cal S}n}^{(i)})$ (top) and $S(\rho^{(2)}_{{\cal S}\bar{\cal S}n})$ (bottom) of the normalized one- and two-body DM blocks in the bosonic case of Fig.\ \ref{fig3}. The bottom panel also depicts that of the normalized collective sub-block \eqref{72b}, $S^{(2)'}_{{\cal S}\bar{\cal S}}=S(\rho^{(2)}_{cn})$. Dashed lines indicate again their maximum values.}    \vspace*{-.35cm}
         \label{fig4}\end{figure}

Figs.\ \ref{fig3}-\ref{fig4}  depict previous quantities in the bosonic case. 
The main difference is the behavior for weak coupling, since  for $g\rightarrow 0^+$ all $m$ pairs now fall to the lowest sp level. This implies a dominant eigenvalue in all blocks $\rho^{(1)}_{\cal S}$, 
$\rho^{(2)}_{\cal S}$ and $\rho^{(2)}_{{\cal S}\bar{\cal S}}$ at low $g$, with $\lambda^{(1)}_k\rightarrow \frac{n}{2}\delta_{k1}$, $\lambda^{(2)}_{kk'}\rightarrow \delta_{kk'}\delta_{k1}\binom{n/2}{2}$   
and $\rho^{(2)}_{ckk''}=\delta_{kk''}\delta_{k1}(\frac{n}{2})^2$ for $g\rightarrow 0$ in \eqref{71}-\eqref{72}. 
As $g$ increases all levels become occupied 
and all eigenvalues of $\rho^{(1)}_{\cal S}$,  $\rho^{(2)}_{\cal S}$ become $<1$ for large $g$, approaching for $g\rightarrow \infty$  the  maximally mixed limits  \eqref{la0}-\eqref{la2} ($\lambda^{(1)}_k\rightarrow \frac{1}{2}$, $\lambda^{(2)}_{kk'}\rightarrow \frac{1}{4}\frac{1-2/n}{1+1/n}\approx\frac{1}{4}(1-\frac{3}{n})$). 
However, in  $\rho^{(2)}_{{\cal S}\bar{\cal S}}$ the dominant eigenvalue $\lambda^{(2)}_1$, though also decreasing  for increasing $g$,  stays well  above $m=n/2$,  approaching \eqref{la1} ($=\frac{3n}{4}-\frac{1}{2}$ for $N=n$)  for $g\rightarrow\infty$. This reflects the strong deviation of the GS  from a permanent as $g$ increases, becoming    approximately a bosonic coboson condensate, where 
a prominent eigenvalue remains  in $\rho^{(2)}$ but not in $\rho^{(1)}$, in contrast with a standard condensate. The paired structure of the bosonic GS for large $g$ can thus be also clearly identified through the spectra of  $\rho^{(2)}$ and $\rho^{(1)}$. 

The associated entropies (using the normalized DM blocks) are depicted in Fig.\ \ref{fig4}. Now they all vanish for $g\rightarrow 0$, while  for $g\rightarrow \infty$ behave as in the fermionic case: those of $\rho^{(1)}_{\cal S}$ and $\rho^{(2)}_{\cal S}$ approach 
 their saturation values  
($S(\rho^{(1)}_{{\cal S}n})\rightarrow \log_2 n$, 
$S(\rho^{(2)}_{{\cal S}n})\rightarrow\log_2\frac{n(n+1)}{2}$), while 
 those of $\rho^{(2)}_{{\cal S}\bar{\cal S}n}$ and the collective sub-block  $\rho^{(2)}_{cn}$ stabilize well below their maximum values ($\log_2 n^2$ and $\log_2 n$, dashed lines),   reaching    the lower limits $\approx \log_2 n^2 -\frac{3}{n}\log_2\frac{3n}{2}$ 
and $\approx\frac{1}{4}\log_2 9.5 n$ (plus $O(n^{-1})$ terms) respectively, reflecting the effect of the remnant dominant eigenvalue. 
\vspace*{-0.5cm}

\subsubsection{Approximate expansions}

     \begin{figure}
     \includegraphics[scale=.349%,trim={6.cm 2.25cm 6cm 2.5cm},clip
     ]{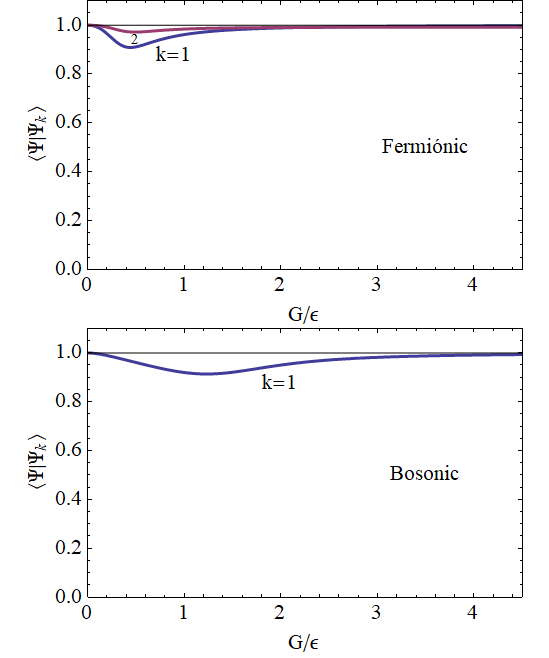}  \vspace*{-0.35cm}
     
     \caption{The overlap $\langle\Psi|\Psi_k\rangle$ between the exact GS $|\Psi\rangle$ and the approximate GS $|\Psi_k\rangle$ obtained by conserving just the first $k$ terms in the $2-(N-2)$ Schmidt-like expansion \eqref{e3b} associated to the central block  $\rho^{(2)}_{{\cal S}\bar{\cal S}}$ in \eqref{rho2ss} of the two-body DM, as a function of the scaled strength $G/\varepsilon$,  in the fermionic (top) and bosonic (bottom) systems of previous figures. Saturation $\langle\Psi|\Psi_k\rangle=1$ (thin line) is always reached for the full expansion associated to this block. In the fermionic case the reduced expansion \eqref{73b} can be used instead of \eqref{e3b}. } 
     \vspace*{-0.5cm}\label{fig5}\end{figure}
 
     We finally depict in  Fig.\ \ref{fig5} the overlap between the exact GS $|\Psi\rangle$ and the approximate normalized GS $|\Psi_k\rangle$ obtained by taking just the first $k$ terms in the Schmidt-like  $(2,N-2)$ expansion of Eq.\ \eqref{e3b}, $|\Psi_k\rangle\propto\sum_{\tilde\nu=1}^k \sigma_{\tilde\nu}^{(2)}A_{\tilde\nu}^{(2)\dag}B_{\tilde\nu}^{(N-2)\dag}|0\rangle$,  based on the two-body DM block $\rho^{(2)}_{{\cal S}\bar{\cal S}}$. Terms are sorted in decreasing order of $\sigma_{\tilde\nu}^{(2)}$, i.e.\ of the eigenvalues $\lambda^{(2)}_\nu=(\sigma^{(2)}_\nu)^2$ of this block. 
     For fermions the expansion can be reduced to the  sum  \eqref{73b} based  on the collective sub-block $\rho^{(2)}_c$, involving just $n$ terms.     
     In particular, for $k=1$ the approximation corresponds to  the dominant eigenvalue $\lambda_1^{(2)}$ and is determined by its normal eigenvector $A_1^{(2)\dag}$: 
     \begin{equation}|\Psi_1\rangle\propto \sigma_{1}^{(2)}\,A^{(2)\dag}_1 B^{(N-2)\dag}_1|0\rangle\,,\label{psi1}\end{equation}
where $\sigma_1^{(2)}B^{(N-2)\dag}_1|0\rangle
=A_1^{(2)}|\Psi\rangle$ (Eq.\ \eqref{17b}). 

          It is first verified that the full sum always yields the exact GS in all expansions. Nonetheless, for those based on $\rho^{(2)}_{{\cal S}\bar{\cal S}}$ (or $\rho^{(2)}_c$  for fermions),  
     already the $k=1$ approximation \eqref{psi1} is seen to provide a very good  overlap  $|\langle\Psi|\Psi_k\rangle|\agt 0.9$  
     for all values of $g$,  minimum just at the transition region around $g\approx 1$.      
     Moreover, this approximation 
     becomes  {exact} for 
 {both} $g\rightarrow\infty$ {\it and} $g\rightarrow 0$ in both the fermionic and bosonic systems, 
     since in these limits the exact GS becomes of the form \eqref{psimp}, i.e.\  Eq.\ \eqref{psi1} with  $B_1^{(N-2)\dag}=(A_1^{(2)\dag})^{m-1}$ and $A_1^{(2)\dag}$ eigenvector of $\rho^{(2)}$: 
     For $g\rightarrow\infty$,   
     $A_1^{(2)\dag}$ becomes  the uniform pair creation operator  \eqref{Ap1}, 
     as the GS approaches the state \eqref{p1}, whereas for
      $g\rightarrow 0$,  
      $A_1^{(2)\dag}\rightarrow c^\dag_1c^\dag_{\bar 1}$ for bosons  while $A_1^{(2)\dag}\rightarrow \frac{1}{\sqrt{m}} \sum_{k=1}^{m} c^\dag_k c^\dag_{\bar k}$ for fermions (as the SD $|\Psi_m^0\rangle$ can be written as $\frac{m^{m/2}}{m!}(A^{(2)\dag}_1)^m|0\rangle$).  
       In the fermionic system the $k=2$ approximation in \eqref{73b} further improves the $k=1$ result in the transition  region, staying also reliable for large $g$,  
      while for bosons more terms are required for obtaining a better approximation for all $g$.   As the terms in these 
 expansions are not necessarily linearly independent, the convergence to the exact $|\Psi\rangle$ is not necessarily monotonous as $k$ increases for all values of $g$. 
 
\section{Conclusions\label{V}}
We have presented a unified formalism for analyzing general  pure states of systems of $N$ indistinguishable particles in terms of exact bipartite-like $(M,N-M)$  decompositions,  valid for both fermions and bosons. It is directly connected with the isospectral $M$ and $N-M$-body DMs, whose eigenvalues acquire here meaning also as coefficients of  the associated diagonal Schmidt decomposition. The ensuing $M$-body entanglement, quantified through the entropy of the normalized $M$-body DM,  was shown to fulfill general monotonicity relations under certain quantum operations.  

We have analyzed in addition the exact reduced expansions emerging when the number of particles in a certain subspace ${\cal S}$ of the full sp space is fixed, which are associated to the ensuing blocks of the DMs. Then both local (in ${\cal S}$) and ``mixed'' (in ${\cal S}$ and its complementary subspace $\bar{\cal S}$) DMs and corresponding  exact  $(M,N\!-\!M)$ expansions arise in connection with these blocks, whose eigenvalues and Schmidt decomposition characterize the system correlations.
The standard reduced DMs and Schmidt decomposition of  distinguishable bipartite systems emerge  as a particular case in this scenario. 

As example, we have analyzed in detail the behavior of the one- and two-body DMs in the GS a finite pairing system. These systems are  characterized by a large dominant eigenvalue of $\rho^{(2)}$ in the superfluid phase, which in the present formalism enables a reliable  description of the exact GS with just a few terms (in fact just one term) of the associated $(2,N-2)$ expansion, in both fermion and boson systems. We have also provided exact results for the eigenvalues of $\rho^{(2)}$ in  the maximally paired states for bosons and fermions, as well as bounds for its dominant eigenvalue  in some general paired states, again for both bosons and fermions. 
Applications to more complex systems and further analysis of the role of present mode-independent entanglement measures in quantum information  are under investigation. 

\acknowledgments 
Authors acknowledge support from CONICET  and CIC (R.R.) of Argentina. Work supported by CONICET PIP Grant No. 11220200101877CO. 

\appendix
\section{Proofs of main expressions\label{A}}
In order to  prove Eq.\ \eqref{Nm}, we first show that  
\begin{equation}
\frac{1}{M!}\sum_{i_1,\ldots,i_M}c^\dag_{i_1}\ldots c^\dag_{i_M}c_{i_M}\ldots c_{i_1}=\binom{\hat{N}}{M}
\label{S1}\end{equation}
for both fermions and bosons, where the sum over each $i_j$ ($j=1,\dots,M$) runs over all $d$ sp  states and $\hat{N}=\sum_i c^\dag_i c_i$ is the particle number operator.

{\it Proof:} When  applying the sum in \eqref{S1} to an arbitrary $N$ particle state $|\Psi\rangle$ with $N\geq M$ (for $N<M$ we set $\binom{N}{M}=0$), the innermost sum  $\sum_{i_m}c^\dag_{i_m}c_{i_m}=\hat{N}$ takes the value  $N-M+1$, and then, successively, the sums  $\sum_{i_j}c^\dag_{i_j} c_{i_j}$ take the values  $N-j+1$ for $j=M,\ldots,1$. This leads to Eq.\ \eqref{S1}.  
Then, using  relations \eqref{cnm}, 
\begin{eqnarray}
\sum_{i_1,\ldots,i_M}\!\!\!\!\tfrac{c^\dag_{i_1}\ldots c^\dag_{i_M}c_{i_M}\ldots c_{i_1}}{M!}
&=&\!\!\sum_{i_1\leq i_2\ldots \leq i_M}\!\!\!\!\!\!\!\!\!\tfrac{c^\dag_{i_1}\ldots c^\dag_{i_M}c_{i_M}\ldots c_{i_1}}{n_1!\ldots n_d!}\nonumber\\=\!\!\!\!\!\!\!\sum_{n_1,\ldots,n_d\atop     n_1+\ldots+n_d=M}\!\!\!\!\!\!\!\!\!\!\tfrac{c^{\dag n_1}_1\ldots c^{\dag n_d}_d c_d^{n_d}\ldots c_1^{n_1}}{n_1!\ldots n_d!}
&=&\sum_{\bm\alpha}C^{(M)\dag}_{\bm\alpha} C^{(M)}_{\bm\alpha}\label{S2}
\end{eqnarray}
where the  sum in  \eqref{S2}  runs over all $d_M$ operators \eqref{Cm}, which leads to Eq.\ \eqref{Nm}. 
Here $n_j$ ($j=1,\ldots,d$) is the number of times sp state $j$ appears in the string $(i_1,\ldots,i_M)$, 
such that $c^\dag_{i_1}\ldots c^\dag_{i_M}=c^{\dag n_1}_1\ldots c^{\dag n_d}_d$, with $n_1+\ldots +n_d=M$. The sum over the ordered  $i_j$'s is equivalent to that over these occupations with previous constraint, with each configuration $(n_1,\ldots,n_d)$ appearing 
$\frac{M!}{n_1!\ldots n_d!}$ times in the first unrestricted sum. These arguments hold for both bosons and fermions, but for the latter are  trivial as $n_i=0,1$.    \qed
 
 We now prove  Eqs.\ \eqref{st2}--\eqref{gamnt}. Using the same previous reasoning, we obtain, for a completely symmetric (bosons) or antisymmetric (fermions) tensor $\Gamma_{i_1,\ldots,i_N}$,  \begin{eqnarray}
\tfrac{1}{N!}\!\!
\!\!\sum_{i_1,\ldots,i_N}\!\!\!\!
\Gamma_{i_1\ldots i_N}\,c^\dag_{i_1}\ldots c^\dag_{i_N}|0\rangle\label{ast1}&=&
\!\!
   \sum_{i_1\leq   \ldots\leq i_N}\!\!\!\!\!\!\Gamma_{i_1\ldots i_N}\tfrac{c^\dag_{i_1}\ldots c^\dag_{i_N}}{n_1!\ldots n_d!}|0\rangle\nonumber\\=\!\!\sum_{n_1,\ldots,n_d\atop \sum_j n_j=N}\!\!\!\!\!\Gamma_{n_1\ldots n_d}\tfrac{c_1^{\dag n_1}\ldots c_d^{\dag n_d}}{n_1!\ldots n_d!}|0\rangle
   &=&\sum_{\bm\alpha} {\Gamma}^{(N)}_{\bm\alpha} C^{(N)\dag}_{\bm\alpha} |0\rangle\,,\label{ast2}
   \end{eqnarray}
 where $\Gamma_{n_1\ldots n_d}=\Gamma_{i_1\ldots i_N}$ if $c^\dag_{i_1}\ldots c^\dag_{i_N}=
 c_1^{\dag n_1}\ldots c_d^{\dag n_d}$
 ($i_1\leq \ldots\leq i_N$ for fermions) and  
 ${\Gamma}^{(N)}_{\bm\alpha}
 =\frac{\Gamma_{n_1\ldots n_d}}{\sqrt{n_1!}\ldots\sqrt{n_d!}}$. \qed 
 
 {\it Relation between $\Gamma^{(M)}_{\bm\alpha\bm\beta}$ and $\Gamma_{i_1\ldots i_N}$}. Rewriting   \eqref{ast1} as  
  \begin{eqnarray}
   |\Psi\rangle&=&\tfrac{M!(N-M)!}{N!}\!\!\!\!\!\!\!\!\!\!\!\sum_{i_1\leq \ldots\leq i_M,\atop i_{M+1}\leq \ldots\leq i_N}\!\!\!\!\!\!\!\!\!\!
   \Gamma_{i_1\ldots i_{M}i_{M+1}\ldots i_N}\tfrac{c^\dag_{i_1}\ldots c^\dag_{i_M}c^\dag_{i_{M+1}}
   \ldots c^\dag_{i_N}}{n_1!\ldots n_d!\,n'_{1}!\ldots n'_d!}|0\rangle\nonumber\\
   &=&{\textstyle\binom{N}{M}^{-1}}\!\!\!\!\!\!\!\!\!\!\!\!\!\!\!\!\!\!\!\!
   \sum_{n_1,\ldots,n_d,n'_1,\ldots,n'_d\atop\sum_j n_j=M,\,\sum_j n'_j=N-M}\!\!\!\!\!\!\!\!\!\!\!\!\!\!\!\!\!\!\!
   \Gamma^{(M)}_{n_1\ldots n_d,n'_1,\ldots,n'_d}\tfrac{c^{\dag n_1}_1\ldots c^{\dag n_d}_d c^{\dag n'_1}_1\ldots c^{\dag n'_d}_d}{n_1!\ldots n_d!\,n'_1!\ldots n'_d!}|0\rangle\nonumber\\
   &=&{\textstyle\binom{N}{M}^{-1}}\sum_{\bm\alpha,\bm\beta}{\Gamma}^{(M)}_{\bm\alpha\bm\beta} C^{(M)\dag}_{\bm\alpha} C^{(N-M)\dag}_{\bm\beta} |0\rangle\,,\label{ast3}
   \end{eqnarray}
 where $\Gamma^{(M)}_{n_1\ldots n_d,n'_1\ldots n'_d}=
 \Gamma_{i_1,\ldots,i_M,i_{M+1}\ldots i_N}$ for $n_j$, $n'_j$ the number of times sp state $j$ appears in  $(i_1,\ldots,i_M)$ and  $(i_{M+1},\ldots,i_N)$  ($i_1<\ldots < i_M$, $i_{M+1}<\ldots<i_N$ for  fermions)  such that $\sum_j n_j=M$, $\sum_j n'_j=N-M$, it is 
 seen that   
 \begin{eqnarray}
 {\Gamma}^{(M)}_{\bm\alpha\bm\beta}&=&
 \frac{\Gamma^{(M)}_{n_1\ldots n_d,n'_1\ldots n'_d}}{\sqrt{n_1!\ldots n_d!}\sqrt{n'_1!\ldots n'_d!}}\label{gammt}\,.\end{eqnarray}  

 {\it Behavior under  unitary sp transformations.} If  \begin{equation} c_i\rightarrow \hat U^\dag c_i \hat U=\sum_k U_{ki} c_k\,,\;\;\;\hat U=e^{-\imath\sum_{i.j} h_{ij}c^\dag_i c_j}\,,\label{Usp}\end{equation} 
 where $h^\dag=h$, $U$ is the matrix $U=\exp[-\imath h]$ and \eqref{Usp} holds for both bosons and fermions,  the operators 
$C^{(M)}_{\bm \alpha}$ transform unitarily: 
 $C^{(M)}_{\bm\alpha}\rightarrow \hat U^\dag C^{(M)}_{\bm \alpha}\hat U=\sum_{\bm \alpha'} U^{(M)}_{\bm \alpha'\bm\alpha} C^{(M)}_{\bm\alpha'}$ for $U^{(M)}$ a unitary symmetrized or antisymmetrized tensor product of $M$ matrices $U$. This implies  $\Gamma^{(M)}\rightarrow U^{(M)*}\Gamma^{(M)}U^{(N-M)\dag}$, 
then leaving its singular values $\sigma^{(M)}_\nu$ unchanged $\forall\,M$. 
 And under similar sp transformations of the state, \begin{equation}|\Psi\rangle\rightarrow \hat U|\Psi\rangle\,,\;\;\;\hat U=e^{-\imath\sum_{i.j} h_{ij}c^\dag_i c_j}\,,\label{Usp2}\end{equation}
we have $\langle \hat{O}\rangle\rightarrow \langle\hat{U}^\dag \hat{O}\hat{U}\rangle$ and hence 
$\rho^{(M)}\rightarrow U^{(M) t}\rho^{(M)}U^{(M)t\dag}$, with $\hat \rho^{(M)}\rightarrow \hat U\hat\rho^{(M)}\hat U^\dag$ as is apparent from   \eqref{rhomo}. Its eigenvalues $\lambda^{(M)}_\nu$  remain then unchanged, in agreement with their direct relation with $\sigma_\nu^{(M)}$. 

{\it Proof of Eqs.\ \eqref{27}--\eqref{res}}:
From Eq.\ \eqref{r1o} we obtain 
${\rm Tr}\, \hat\rho^{(M)}C^{(L)\dag}_{\bm\gamma'}C^{(L)}_{\bm\gamma}\!\!\!=$ {\small $\sum_{\bm\beta}\langle \Psi|C^{(N-M)\dag}_{\bm\beta}C^{(L)\dag}_{\bm\gamma'}C^{(L)}_{\bm\gamma}C_{\bm\beta}^{(N-M)}|\Psi\rangle$} $=\binom{N-L}{N-M}\rho^{(L)}_{\bm \gamma\bm\gamma'}$ for $L\leq M\leq N$,  by using Eq.\ \eqref{Nm} for $\sum_{\bm\beta}C^{(N-M)\dag}_{\bm\beta}C^{(N-M)}_{\bm \beta}$ applied on the  $N-L$-particle state $C_{\bm\gamma}^{(L)}|\Psi\rangle$, 
with $\rho^{(L)}_{\bm\gamma\bm\gamma'}=\langle\Psi|
C^{(L)\dag}_{\bm\gamma'}C^{(L)}_{\bm\gamma}|\Psi\rangle$. This  leads to Eq.\  \eqref{27}. 
Then, by replacing expression \eqref{r1o} or \eqref{31} in \eqref{27}, we obtain, 
for the normalized DM $\rho^{(L)}_n=\rho^{(L)}/\binom{N}{L}$, 
\begin{eqnarray}
\rho_n^{(L)}&=&\tfrac{\binom{N}{M}\binom{M}{L}}{\binom{N}{L}\binom{N-L}{N-M}}\sum_{\bm \beta} p_{\bm\beta}\,\rho_{\bm\beta n}^{(L)}
\,,\label{A8}\end{eqnarray}
where $\rho_{\bm\beta n}^{(L)}=\rho_{\bm\beta}^{(L)}/\binom{M}{L}$ are the normalized $L$-body DMs in the normalized $M$-particle states $|\Psi_{\bm\beta}\rangle=\frac{1}{\sqrt{p_{\bm\beta}}}{\cal M}_{\bm\beta}|\Psi\rangle$, with $p_{\bm\beta}$ the probabilities \eqref{pbet}. This leads to Eq.\ \eqref{res}
since $\binom{N}{M}\binom{M}{L}/[\binom{N}{L}\binom{N-L}{N-M}]=1$, 
in agreement with normalization: ${\rm Tr}\,\rho_n^{(L)}={\rm Tr}\,\rho_{\bm\beta n}^{(L)}=1=\sum_{\bm\beta}p_{\bm\beta}$. \qed

We now prove the contractions and results \eqref{la0},\eqref{la2} and \eqref{70} in the maximally paired state \eqref{58} for bosons and fermions. The number of states with $m$ pairs $(k,\bar k)$ in a sp space of $n$ sp states $k$ and $n$ sp states $\bar{k}$ is 
\begin{eqnarray}
{\cal N}_{n,m}&=&\left\{\begin{array}{ccl}
\binom{n+m-1}{m}&&{\rm (bosons)}\\ &&\vspace*{-0.25cm}\\
\binom{n}{m}&&{\rm (fermions)}\end{array}\right.\,.
\end{eqnarray}
where $m\geq 0$ for bosons and $0\leq m\leq n$ for fermions. 
Then, for averages $\langle O\rangle=\langle\Psi_m|O|\Psi_m\rangle$ in the state \eqref{58}, we obtain in the first place the expected obvious result \begin{eqnarray}
\langle c^\dag_k c_{k'}\rangle&=&\delta_{kk'}
{\textstyle\sum_{l} \frac{l\,{\cal N}_{n-1,m-l}}{{\cal N}_{n,m}}=\frac{m}{n}}
\end{eqnarray}
for both bosons and fermions, where the sum runs over $l=0,1,2,\ldots,m$ for bosons and  $l=0,1$ for fermions. The same holds for $k\rightarrow \bar k$. 
For two body contractions, assuming in what follows $k\neq k'$, we obtain 
\begin{eqnarray}
\langle c^\dag_k\,c^\dag_{k'} c_{k'}c_k\rangle\!&=&
\!\!\!{\textstyle\sum_{l,l'}\frac{ll'{\cal N}_{n-2,m-l-l'}}{{\cal N}_{n,m}}=\frac{m(m\!-\!1)}{n(n\!\pm\! 1)}=\lambda^{(2)}_2}\;\;\;\;\;\;\;\;\;\;\;\label{la2ap}
\end{eqnarray}
 where the sum runs over $l+l'\leq m$ for bosons ($+$) and $l=l'=1$ for fermions ($-$). This same result is obtained for 
$\langle c^\dag_k c^\dag_{\bar k'}c_{\bar k'} c_k\rangle$,  $\langle \frac{c^{\dag\,2}_k c_k^2}{2!}\rangle$ (boson case) and $k,k'\rightarrow \bar k,\bar k'$. 
On the other hand, 
\begin{eqnarray}
\langle c^\dag_k\,c^\dag_{\bar k} c_{\bar k}c_k\rangle\!&=&
\!\!{\textstyle\sum_{l}\frac{l^2\,{\cal N}_{n-1,m -l}}{{\cal N}_{n,m}}=\frac{m(n\!+\!m\!-\!1\!\pm\! m)}{n(n\!\pm\!1)}}
\label{rhopd}\;\;\;\;\;\;\\
\langle c^\dag_k\,c^\dag_{\bar k} c_{\bar k'}c_{k'}\rangle\!&=&
\!\!{\textstyle\sum_{l,l'}\frac{l(l'\!+\!1)\,{\cal N}_{n-2,m -l-l'}}{{\cal N}_{n,m}}=\frac{m(n\!\pm\! m)}{n(n\!\pm\!1)}}.\;\;\;\;\;\;\;\;\label{rhopo}
\end{eqnarray}
These exact results lead to Eqs.\ \eqref{la0}--\eqref{70}. 

Then blocks $\rho^{(2)}_{\cal S}$ and $\rho^{(2)}_{\bar{\cal S}}$ of $\rho^{(2)}$ become here identical and  proportional to 
$\lambda^{(2)}_2\mathbbm 1$, whereas the mixed block $\rho^{(2)}_{{\cal S}\bar{\cal S}}$ becomes itself blocked in two submatrices:
\begin{equation}
\rho^{(2)}_{{\cal S}\bar{\cal S}}=
\begin{pmatrix}\lambda^{(2)}_2\mathbbm 1&0\\0&\rho^{(2)}_{c}\end{pmatrix}\label{pmatrix}\,,\end{equation}
where $\rho^{(2)}_{c_{kk'}}=\langle c^\dag_k c^\dag_{\bar k}c_{{\bar k}'}c_{k'}\rangle=a\delta_{kk'}+b(1-\delta_{kk'})$ is the $n\times n$ block containing the two-body pairing correlations, with $a,b$ given by \eqref{rhopd}--\eqref{rhopo}. 
Its eigenvalues are then $\lambda^{(2)}_1=(n-1)b+a$, Eq.\ \eqref{la1}, non-degenerate and dominant, and  $\lambda^{(2)}_2=a-b$,  Eqs.\ \eqref{la2}--\eqref{la2ap}, $n-1$-fold degenerate. \qed

Proof of Eq.\ \eqref{Fg}: For bosons, 
 a direct evaluation of  
 $\langle A^\dag A\rangle$  
in the state \eqref{psimp}, with $A$ given by \eqref{Ad},  yields   
\begin{equation}
\begin{split}
\langle A^\dag A\rangle&=
\sum_k\sigma_k^2 \langle m^2_k\rangle+\sum_{k\neq k'}\sigma_k^2\langle(m_k+1)m_{k'}\rangle\label{AF1}\\ 
&=m+(m-1)\sum_k\sigma_k^2\langle c^\dag_k c_k\rangle 
\end{split}
\end{equation}
where we used $\sum_k\sigma_k^2=1$,  $\sum_k m_k=m$ and $\langle m_k\rangle=\langle c^\dag_k c_k\rangle=\langle c^\dag_{\bar k}c_{\bar k}\rangle$.  Similarly, for fermions we obtain  
\begin{equation}\begin{split}
\langle A^\dag A\rangle&=
\sum_k \sigma_k^2\langle m_k\rangle+\sum_{k\neq k'}\sigma_k^2\langle(1-m_k)m_{k'}\rangle\\ 
&=m-(m-1)\sum_k\sigma_k^2\langle c^\dag_k c_k\rangle\,.
\label{AF2}
\end{split}
\end{equation}
Eqs.\ \eqref{AF1}--\eqref{AF2} then lead to Eq.\ \eqref{Fg}. \qed
\vspace*{-0.5cm}

 \section{$M$-body entanglement and bounds to bipartite entanglement after particle transfer\label{B}}
  Let us consider a general  initial state   (Eq.\ \eqref{Psm}) \begin{equation}|\Psi_0\rangle={\textstyle\binom{N}{M}^{-1}}\sum_{\bm\alpha\bm\beta\in{\cal S}_0}\Gamma^{(M)}_{\bm\alpha,\bm\beta}C^{(M)\dag}_{\bm\alpha}C^{(N-M)\dag}_{\bm\beta}|0\rangle\end{equation}  of $N$ indistinguishable particles (fermions or bosons), occupying sp states just  within a subspace ${\cal S}_0\subset{\cal H}$ of the full sp space ${\cal H}$.  
   We then consider a completely positive trace preserving (CPTP) operation ${\cal T}$  which  transfers  $M<N$ particles from ${\cal S}_0$ to an initially empty subspace ${\cal S}$ orthogonal to ${\cal S}_0$ (e.g., $M$ particles from a group of $N$ localized within  some bounded region of space to a distinct non overlapping region, or from 
    low lying energy levels to higher orthogonal levels). For this purpose we define the $M$-body operators $\hat  T_r$   
  such that 
  \begin{subequations}
  \label{57}
 \begin{eqnarray} \hat{T}_r|\Psi_0\rangle&=&{\textstyle\binom{N}{M}^{-1/2}}\!\!\!\!\!\sum_{\bm\gamma\in{\cal S},\bm\alpha\in{\cal S}_0}\!\!T_{r_{\bm\gamma\bm\alpha}}C^{(M)\dag}_{\bm\gamma}C_{\bm\alpha}^{(M)}|\Psi_0\rangle\label{57a}\;\;\;\;\\&=&\sum_{\bm\gamma\in{\cal S},\bm\beta\in{\cal S}_0}\!\!\Gamma^{(M)}_{r_{\bm\gamma\bm\beta}}C^{(M)\dag}_{\bm\gamma}C^{(N-M)\dag}_{\bm\beta}|0\rangle\,,\label{stdis}\end{eqnarray}
 \end{subequations}
is a state satisfying \eqref{Ns} for $N_{\cal S}=M$, where 
$\Gamma^{(M)}_r=\binom{N}{M}^{-1/2} T_r\Gamma^{(M)}$ 
(Eq.\ \eqref{Cbet}). 
If we now assume 
$\sum_r T_r^\dag T_r=\mathbbm{1}$ and use Eq.\ \eqref{Nm} for $N\rightarrow M$,  then for any such $|\Psi_0\rangle$, 
\begin{equation}{\textstyle\sum_r} \hat T_r^\dag\hat T_r|
\Psi_0\rangle=|\Psi_0\rangle\,,\end{equation} 
since $C_{\bm\gamma'}^{(M)}C^{(M)\dag}_{\bm\gamma}C^{(M)}_{\bm\alpha}|\Psi_0\rangle=\delta_{\bm\gamma'\bm\gamma}C^{(M)}_{\bm\alpha}|\Psi_0\rangle$ for $\bm\gamma',\bm\gamma\in{\cal S}$, $\bm\alpha\in {\cal S}_0$, for both fermions and bosons. Hence we can consider the set of operators $\hat T_r$ as a quantum operation mapping $|\Psi_0\rangle$ to states $|\Psi_r\rangle\propto 
\hat T_r|\Psi\rangle$ with probabilities \begin{equation}
p_r=\langle \Psi_0|\hat T_r^\dag\hat T_r|\Psi_0\rangle={\rm Tr}[\Gamma^{(M)\dag}_r\Gamma^{(M)}_r]
\,,\end{equation}  
satisfying $\sum_r p_r
= {\rm Tr}\,[\Gamma^{(M)\dag}\Gamma^{(M)}]/\binom{N}{M}=1$. 

Using  result \eqref{rhons}, the reduced state of the $N_{{\cal S}}=M$ particles at ${\cal S}$ in the normalized state $|\Psi_r\rangle$ is 
$\rho^{(M)}_{{\cal S}r}=\Gamma^{(M)}_r\Gamma^{(M)\dag}_r/p_r$. Since the nonzero eigenvalues of $\Gamma^{(M)}_r\Gamma^{(M)\dag}_r$ are the same as those of $\Gamma^{(M)\dag}_r\Gamma^{(M)}_r$ and since  $\sum_r \Gamma^{(M)\dag}_r\Gamma_r^{(M)}=\Gamma^{(M)\dag}\Gamma^{(M)}/\binom{N}{M}$ has then the same nonzero eigenvalues as the normalized DM $\rho^{(M)}_{0n}=\Gamma^{(M)}\Gamma^{(M)\dag}/\binom{N}{M}$ in the original state $|\Psi_0\rangle$,  we obtain the following majorization relation   
\begin{equation} \bm{\lambda}(\hat\rho_{0n}^{(M)})\prec \sum_{r}p_{r}\bm{\lambda}(\hat\rho^{(M)}_{{\cal S}r})\label{prec2}\end{equation}
between the sorted eigenvalues of $\rho^{(M)}_{0n}$ and those of $\rho^{(M)}_{{\cal S}r}$, the latter determining the entanglement between the $M=N_{\cal S}$ particles at ${\cal S}$ and the remaining $N-M$ particles at the orthogonal subspace ${\cal S}_0$. It implies the following entropic inequality
\begin{equation}S(\hat\rho_{0n}^{(M)})\geq \sum_{r}p_r S(\hat\rho^{(M)}_{{\cal S}r})\label{sent}\,,
\end{equation}
between the entropy of the normalized DM  $\rho^{(M)}_{0n}$ in $|\Psi_0\rangle$, which measures its $M$-body entanglement, and the average entanglement entropy between the (now distinguishable) $M$ particles at ${\cal S}$ and the remaining $N-M$ particles at ${\cal S}_0$ in the post-selected states $|\Psi_r\rangle$, which then cannot surpass the original $M$-body entropy.
It is valid again for any concave concave entropy $S$. All other results derived in  \cite{GDR.21} for the conversion $|\Psi\rangle\rightarrow \{|\Psi_r\rangle\}$ in fermion systems remain then valid for bosons. 

Finally, note that if $T_{r_{\bm\gamma\bm\alpha}}=T_{\bm\gamma\bm\alpha}\delta_{r\bm\alpha}$ ($\bm\alpha\in{\cal S}_0$),  then    \begin{equation}\hat T_{\bm\alpha}|\Psi_0\rangle={\textstyle\binom{N}{M}^{-1/2}
\sum_{\bm\gamma\in{\cal S}}}T_{\bm\gamma\bm\alpha}C^{(M)\dag}_{\bm\gamma}C^{(M)}_{\bm\alpha}|\Psi_0\rangle\,,\label{B7}\end{equation} with $\sum_{\bm\gamma}|T_{\bm\gamma\bm\alpha}|^2=1$ and hence $\sum_{\bm\alpha}\hat T_{\bm\alpha}^\dag\hat T_{\bm\alpha}|\Psi_0\rangle=|\Psi_0\rangle$.  Thus, this map implements the measurement based on the operators \eqref{Kr} (for $N-M\rightarrow M$)  on ${\cal S}_0$  through a particle number conserving map in the full system ${\cal S}_0\oplus {\cal S}$, transferring   $M$ particles from  ${\cal S}_0$ 
to ${\cal S}$. 

%\bibliography{bibtex.bib}

\end{document}